\documentclass[a4paper,11pt]{article}

\usepackage{jcappub} 

\usepackage[T1]{fontenc} 
\usepackage{graphicx}
\usepackage{dcolumn}
\usepackage{amsfonts}

\usepackage{amssymb}
\usepackage{amsmath}
\usepackage{graphicx}
\usepackage{subfigure}
\usepackage{makeidx}
\usepackage{capt-of}
\usepackage{float}
\usepackage{placeins}
\usepackage{perpage}
\usepackage{mwe}
\usepackage{kantlipsum}
\graphicspath{{images/}}

\title{\Large{Constraining chameleon field driven  warm inflation with Planck 2018 data}}

\author[a,b,1]{Haidar Sheikhahmadi,\note{Corresponding author.}}
\author[c,d]{Abolhassan Mohammadi,}
\author[e]{Ali Aghamohammadi,}
\author[f,g]{Tiberiu Harko,}
\author[h]{Ram\'{o}n Herrera,}
\author[i,j]{Christian Corda,}
\author[a]{Amare Abebe,}
\author[c]{Khaled Saaidi}

\affiliation[a]{Center for Space Research, North-West University, Mafikeng, South Africa}
\affiliation[b]{School of Astronomy, Institute for Research in
Fundamental Sciences (IPM),  P. O. Box 19395-5531, Tehran, Iran}
\affiliation[c]{Department of Physics, Faculty of Science, University of Kurdistan,  Sanandaj, Iran}
\affiliation[d]{Dipartimento di Fisica e Astronomia, Università di Bologna, via Irnerio 46, I-40126 Bologna, Italy}
\affiliation[e]{Sanandaj Branch, Islamic Azad University, Sanandaj, Iran}
\affiliation[f]{Department of Physics, Babes-Bolyai University, Kogalniceanu Street,
Cluj-Napoca 400084, Romania}
\affiliation[g]{School of Physics, Sun Yat-Sen University, Xingang Road, Guangzhou 510275, People's
Republic of China}
\affiliation[h]{Instituto de F\'{\i}sica, Pontificia Universidad de Cat\'olica de
Valpara\'{\i}so,
Casilla 4950, Valpara\'{\i}so, Chile}
\affiliation[i]{Research Institute for Astronomy and
Astrophysics of Maragha (RIAAM), P.O. Box 55134-441, Maragha, Iran}
\affiliation[j]{International Institute for Applicable Mathematics \& Information
Sciences (IIAMIS),  B.M. Birla Science Centre, Adarsh Nagar, Hyderabad
- 500 463, India}
\emailAdd{h.sh.ahmadi@gmail.com;h.sheikhahmadi@ipm.ir}
\emailAdd{abolhassanm@gmail.com;a.mohammadi@uok.ac.ir}
\emailAdd{a.aqamohamadi@gmail.com}
\emailAdd{tiberiu.harko@gmail.com}
\emailAdd{ramon.herrera@pucv.cl}
\emailAdd{cordac.galilei@gmail.com}
\emailAdd{amare.abbebe@gmail.com}
\emailAdd{khaledsaeidi@gmail.com}

\abstract{We investigate warm inflationary scenario in which the accelerated expansion of the early Universe is driven by chameleon-like scalar fields.  Due to the non-minimal coupling between the
scalar field and the matter sector, the energy-momentum tensor of each fluid component is not conserved anymore, and the generalized balance equation is obtained. The new
source term in the energy equation can be used to model warm inflation. On the other hand,  if the
coupling function varies slowly, the model reduces to the standard model used for the description of cold inflation. To
test the validity of the warm chameleon inflation model, the results for  warm inflationary scenarios are compared
with the observational Planck2018 Cosmic Microwave Background data. In this regard, the perturbation parameters such as the amplitude
of scalar perturbations, the scalar spectral index and the tensor-to-scalar ratio are derived
at the horizon crossing in two approximations, corresponding to the weak and strong dissipative regimes. As a general result it turns out that the theoretical predictions of the chameleon warm inflationary scenario are consistent with the Planck 2018 observations.\\
\\
\textbf{pacs}:~04.20.-q, 04.20.CV, 04.20.Dw, 04.25.dc\\
\\
\textit{keywords:} Warm inflation,  Chameleon mechanism, dissipation function, Planck 2018}
\usepackage{mathrsfs}
\begin{document}
\maketitle
\flushbottom
\newpage

\newcommand{\ber}{\begin{eqnarray}}
\newcommand{\eer}{\end{eqnarray}}
\newcommand{\bern}{\begin{eqnarray*}}
\newcommand{\eern}{\end{eqnarray*}}
\newcommand{\beast}{\begin{equation*}}
\newcommand{\eeast}{\end{equation*}}
\newcommand{\nn}{\nonumber\\}
\newcommand{\bw}{\begin{widetext}}
\newcommand{\ew}{\end{widetext}}
\def\be{\begin{equation}}
  \def\ee{\end{equation}}
\def\bea{\begin{eqnarray}}
\def\eea{\end{eqnarray}}
\def\f{\frac}
\def\n{\nonumber}
\def\l{\label}
\def\p{\phi}
\def\o{\over}
\def\R{\rho}
\def\pa{\partial}
\def\om{\omega}
\def\na{\nabla}
\def\P{\Phi}

\section{Introduction}\label{Intro}

Four decades after the introduction of the inflation model, it can now be considered as one of the cornerstones of modern cosmology \cite{Guth:1980zm,Linde:1981mu,Albrecht:1982wi,Linde:1983gd,Linde:2007fr,Kazanas:1980tx,Martin:2003bt,Martin:2004um,Martin:2007bw}. The main success of this theory goes back to resolving the three main drawbacks of the Standard Big Bang (SBB) theory, namely the flatness, horizon and relic problems \cite{Starobinsky:1979ty,Mukhanov:1981xt, Hawking:1982cz, Starobinsky:1982ee, Guth:1982ec, Bardeen:1983qw}. In inflation theory we deal with very high energy scales, between $200$ GeV and $10^{12}$ TeV \cite{Martin:2013tda}, and, by considering the quantum effects inherent to these energies, one can explain the origin of seeds for Large-Scale Structure (LSS) formation, besides the fluctuations generating the Cosmic Microwave Background (CMB) anisotropies \cite{Martin:2013tda,Stewart:1993bc,Mukhanov:1990me, Liddle:2000cg,Liddle:1994dx,Bennett:2012fp,Hinshaw:2012fq,Wang:1999vf, Gangui:1993tt,Gangui:1994yr}. In the present work, we will investigate in more detail the relation between quantum fluctuations and classical behaviour at the end of inflation.

Inflationary scenarios usually predict that the power spectrum originated from primary quantum fluctuations should have equal power on all scales, {\it i.e.}, it should be scale invariant \cite{Kiefer:1998qe,
  Martin:2004um, Martin:2007bw, Kiefer:2008ku, Sudarsky:2009za,
  Martin:2012pea, Martin:2012ua}. Although  several  interesting models have been proposed to explain the origin of the Universe \cite{Alexander:2000xv,
  Steinhardt:2001st, Khoury:2001bz, Khoury:2001wf, Martin:2001ue,
  Steinhardt:2002ih, Finelli:2001sr, Brandenberger:2001kj,
  Kallosh:2001ai, Martin:2002ar, Peter:2002cn, Tsujikawa:2002qc,
  Kofman:2002cj, Khoury:2003rt, Martin:2003bp, Martin:2003sf,
  Martin:2004pm, Nayeri:2005ck, Peter:2006hx, Finelli:2007tr,
  Abramo:2007mp, Falciano:2008gt, Linde:2009mc, Abramo:2009qk,
  Brandenberger:2009yt, Brandenberger:2011et, Brandenberger:2012zb,
  Cai:2012va, Cai:2013vm,Saaidi:2012qp}, inflation, due to its potential for solving the major problems of the SBB theory, has become a crucial part of present day cosmology. To generate an inflationary phase for the very early evolution of the Universe, which could explain the symmetries of the cosmological principle, one of the important physical candidates is a scalar field. This type of exotic matter can be described as a fluid with negative pressure that inflation needs to have. In other words, the logarithm of the scalar field potential should move much slower than the kinetic part,  leading to the rapid exponential expansion determining the inflationary expansion \cite{Guth:1980zm,Linde:1981mu,Albrecht:1982wi,Linde:1983gd,Linde:2007fr,Kazanas:1980tx}.

  Despite all of its  successes the nature and the origin of the inflationary  theory remain a mystery, and it is still a matter of intense scientific investigation \cite{Starobinsky:1979ty,Mukhanov:1981xt, Martin:2013tda}. Nevertheless, this is not so surprising for us because inflation happened, as mentioned before, in a very high energy epoch. Up to now, there is no complete or convincing version of the Grand Unification Theory (GUT) that could describe the physics of the very early Universe, and of its beginning.  Following \cite{Martin:2013tda}, and from the observational point of view,  we can classify the huge number of different models of inflation into three main categories, with their related sub-categories. These are the single-field inflation \cite{Arkani-Hamed:2015bza,Dimastrogiovanni:2015pla,Chen:2015lza,Chen:2016cbe,Lee:2016vti,Chen:2016qce,Meerburg:2016zdz,Chen:2016uwp,Chen:2016hrz,An:2017hlx,An:2017rwo,Iyer:2017qzw,
  ArmendarizPicon:1999rj,Sheikhahmadi:2015gaa}, the multiple-field inflation \cite{Bassett:2005xm,Wands:2007bd,Emami:2013lma,Sheikhahmadi:2016wyz,Sheikhahmadi:2019xkx,Alishahiha:2004eh, Langlois:2008qf, Langlois:2009ej} and those models in which the fluid is not described by scalar fields \cite{Golovnev:2008cf,Adshead:2012kp,Maleknejad:2011jw,Maleknejad:2011sq,Maleknejad:2012fw}.

One of the important requirements of the inflationary scenarios is to establish a physically realistic relation between the end of the early accelerated, de Sitter type evolution, and the beginning of the radiation-dominated epoch, which corresponds to the SBB model. Based on this requirement, one can consider two different types of inflation,  namely the super-cold and the warm models of inflation\cite{Berera:1995ie,Berera:1996nv,Berera:1995wh,Berera:1996fm,Berera:1998gx,Berera:1999ws,
Bastero-Gil:2016qru,Berera:2018tfc,Yokoyama:1998ju,BasteroGil:2010pb,Bartrum:2013oka,Naderi:2018kre,Ghadiri:2018nok}.

Whereas each one of these scenarios has its advantages and drawbacks, in the present work we want to consider both of them in more detail for a single-field model of inflation, based on the chameleon mechanism. But before going through this model, let us explain the properties of super-cold and warm proposals of inflation and their successes and failures. In super-cold models, to establish a relation between the end of inflation and  the beginning of the hot big bang model, usually the concept of quantum fluctuations, based on the scalar field oscillations, is considered. In other words, in this approach, one should assume a tachyonic pre-heating phase. Hence, based on this mechanism, the end of inflation is smoothly connected to the radiation-d
dominated epoch \cite{Turner:1983he,
  Kofman:1997yn, Bassett:2005xm, Mazumdar:2010sa, Finelli:1998bu,
  Bassett:1998wg, Finelli:2000ya,Jedamzik:2010dq, Jedamzik:2010hq,
  Easther:2010mr}.   The main problem of this model goes back to the consistency with observations. Unfortunately we know very little, from an observational point of view, about these eras and their evolutions, {\it i.e.}, the pre-heating and the reheating, and we have no accurate criteria to compare the theoretical results with observations  \cite{ Martin:2013tda}. Recently, different proposals have been proposed to bind existing observational data together  to obtain a better estimation of super-cold inflation \cite{Ade:2013ktc,Bennett:2012fp, Hinshaw:2012fq,Ade:2013uln, Ade:2013ydc,Tonry:2003zg, Riess:2004nr, Riess:2006fw, Riess:2011yx,AdelmanMcCarthy:2007aa, Abazajian:2008wr,Amiaux:2012bt,Turner:1993vb, Maggiore:1999vm,
  Kudoh:2005as, Kuroyanagi:2009br, Kawamura:2011zz,
  AmaroSeoane:2012km, Kuroyanagi:2013ns,Dunkley:2013vu, Sievers:2013wk,Hou:2012xq, Story:2012wx,Baumann:2008aq, Crill:2008rd,Zaldarriaga:2003du, Lewis:2007kz,
  Tegmark:2008au, Barger:2008ii, Mao:2008ug, Adshead:2010mc,
  Clesse:2012th}.

  Another important constraint originating from this setup, besides the adiabatic initial condition on the CMB, is that the temperature due to reheating has to be larger than the BB Nucleosynthesis (BBN) scales. For more details and for different solutions to overcome this drawback,  we refer the reader to Ref. \cite{Martin:2010kz}, and references therein.

In the present paper we are going to investigate the properties of warm inflation in more detail. A necessary condition for a  field theory to produce warm inflation is that the radiation must be in thermal equilibrium. A crucial advantage of this model of inflation is that it does not need any  preheating and reheating mechanisms to realize the connection between the end of inflation and the beginning of the radiation era \cite{Berera:1995ie,Berera:1996nv,Berera:1995wh,Berera:1996fm,Berera:1998gx}.  To see how this can be achieved, let us go back to the aforementioned crucial problem of initial quantum fluctuations, related to the super-cold inflation. As opposed to the super-cold model of inflation, in warm inflation one can assume the existence of an interaction  between the different components of the initial stages of the Universe  as an intrinsic physical property. This assumption plays the role of a master-key for warm inflation. In other words, when the scalar field acquires the zero-point energy of the inflaton, the responsible field in driving inflation, its reaction back on the inflaton can cause the damping of its motion \cite{Berera:1998gx,Berera:1999ws,Sheikhahmadi:2014rka}. We must stress that to solve the flatness and horizon problems this reaction should be strong \cite{Berera:1998gx,Berera:1999ws}. Therefore instead of slow-roll mechanisms  in warm inflation the concept of over-damped motion can be introduced \cite{Berera:1998gx,Berera:1999ws}.

The prediction of the cosmological quantum perturbations is one of the main achievements of the inflationary scenario. The perturbations are divided into three types: scalar, vector, and tensor perturbations, which evolve independently up to the linear order. The scalar perturbations are known to be the seeds of the LSS of the Universe, while the tensor perturbations are primordial gravitational waves.
 For the sake of completeness, we recall that the now-famous event GW150914,
which is the first direct observation of gravitational waves from
a binary black hole merger \cite{key-1} occurred in the 100th anniversary
of Albert Einstein's prediction of gravitational waves \cite{key-2}.
That event was a cornerstone for science in general and for gravitational
physics in particular. It indeed gave definitive proof of the existence
of gravitational waves, of the existence of black holes having mass greater
than 25 solar masses, and of the existence of binary systems of black
holes that merge in a time less than the age of the Universe \cite{key-1}.

Such a direct gravitational waves detection represented the starting point
of the new era of gravitational waves astronomy. After the event GW150914, the LIGO Scientific Collaboration
announced other six new gravitational waves detections, the events
GW151226 \cite{key-3}, GW170104 \cite{key-4}, GW170814 \cite{key-5},
GW170817 \cite{key-6}, GW170608 \cite{key-7} and, recently, GW151012
\cite{key-8}. All the cited events again are the consequences of binary black
hole coalescences, with the sole exception of the event GW170817, which
represents the first gravitational waves detection from a neutron star
merger \cite{key-6}. Gravitational waves detectors could be, in principle,
decisive to confirm the physical consistency of Einstein's general
theory of relativity, or, alternatively, to endorse the framework
of extended theories of gravity \cite{key-9,key-10}. In fact, some
differences between the GTR and alternative theories appear
in the linearized theory of gravity, and they could be observed through different interferometer
response functions \cite{key-9,key-10,key-11}.

As mentioned already, important candidates for the initial seeds of LSS formation are the quantum fluctuations. But the main question is how we can find a mechanism to link quantum fluctuations to these classical seeds, namely the quantum-to-classical transition problem \cite{ Guth:1985ya,Lyth:1984gv, Halliwell:1986ja,Sasaki:1986hm}. In \cite{Sasaki:1986hm} it was shown that although by considering modes larger than the Hubble length one can treat the fluctuations classically, they are not covariant under canonical transformation \cite{Guth:1985ya,Lyth:1984gv}. To solve this problem,  a solution was proposed  in \cite{Halliwell:1989vw,Calzetta:1990bb, Paz:1991ze,Calzetta:1993qe, Matacz:1992tp}, which is based on the neglect of the quantum interaction between macroscopically distinguished events, namely decoherence. Whereas in warm inflation the dissipation behaviour appears as a basic property, the decoherence can automatically generate it.  In  \cite{Berera:1998gx,Berera:1998px,Kubo:1965,Kubo:1957mj,Callen:1951vq,Caldeira:1981rx} different conditions for warm inflation originating from quantum field theory  were studied. \\

The present paper is organized as follows. We first briefly review the chameleon scalar field model in Section~\ref{SecIIa}. In particular, we introduce the gravitational action, involving a non-minimal coupling between scalar field and the matter sector, and we write down the generalized Friedmann equations of the theory, as well as the energy balance equation, in which the coupling between scalar field and matter generates a new source term, given by the product of the derivative of the coupling function with respect to the scalar field, and the matter Lagrangian. The chameleon warm inflation model is introduced, and analyzed, in Section~\ref{SecIV}. First, in Section~\ref{SecIVa} we review the standard approach to warm inflation, in which a radiation component is present, together with the scalar, during inflation. Due to the interaction between these two components, there is an energy transfer from the scalar field to radiation, due to the presence of a new source term in the conservation equations. The slow-roll parameters and the number of e-folds are also obtained. The main theoretical model we investigate in the paper, the chameleon field driven warm inflation model, is presented in detail in Section~\ref{warmchameleon}. In the chameleon warm inflationary model the dissipative term in the conservation equations is automatically generated due to the scalar field-radiation coupling. This also fixes the functional forms of the Hubble function, of the scalar field potential, and of the radiation temperature as functions of the coupling function $f(\phi)$, and, consequently, of the scalar field. Under the assumption that the temperature dependence of the radiation is a power law function of the scalar field, the temperature evolution can be fully determined. Moreover, the scalar spectral index, the amplitude of tensor perturbations, and the tensor spectral index, are also obtained. Usually, the warm inflationary models are investigated in two limiting regimes, corresponding to weak and strong dissipation, respectively. In Section~\ref{SecVa} we investigate the weak dissipative regime, for which the dissipative ratio is much smaller then unity. We perform a detailed physical and numerical analysis of the model in this approximation, by obtaining the slow-roll parameters, the scalar spectral index $n_s$, and the tensor to scalar ratio $r$ as functions of the coefficients $n$ and $m$ (characterizing the temperature evolution), and of the number of e-folds. Using the $r-n_s$ diagram from the Planck 2018 data, the comparison of the theoretical predictions of the model with the observational data is performed, and the allowable range of parameters is obtained. The evolution of the scalar field potential and of the coupling function $f(\phi)$ is also obtained.  The strong dissipative regime is investigated in a similar way in Section~\ref{SecVb}, and the allowable ranges of the model parameters are obtained from the comparison with the observational data. The coupling function has an exponential behavior. Finally, in Section~\ref{conclusion} we discuss and conclude our results.


\section{A brief review of the chameleon model}\label{SecIIa}

As stated earlier, a scalar field-driven early epoch of inflationary expansion is nowadays taken to be as a standard ingredient to describe the expansionary history of the Universe.
 Less standard, however, is the nature of the inflationary scenario, and of the driving scalar field. One possible candidate suggested in the literature \cite{Mota:2003tc,Khoury:2003aq,Khoury:2003rn,Brax:2004px,Khoury:2013yya,Waterhouse:2006wv,Clifton:2006vm,
 Das:2008iq,Saaidi:2011zza,Saaidi:2012rj,Saaidi:2010ts,Farajollahi:2010bn,Farajollahi:2011ym} is the so-called chameleon scalar field, a light scalar field whose mass depends on the ambient matter density. Originally put forward to overcome the quintessence mechanism drawbacks, and as a dark energy candidate for the late-time cosmic acceleration, recent studies  \cite{ Mota:2011nh,Hinterbichler:2013we, Creminelli:2013nua,Saba:2017xur} have shown that chameleonic inflation is also possible with the appropriate choice of the potential for the scalar field. This leads to a varying mass for the scalar field in which in a dense environment, the scalar field acquires a large mass, which results in short range effects.

In the original chameleon model, the scalar field is coupled to the matter through a conformal factor, $\tilde{g}_{\mu\nu}= e^{2\beta \phi \over M_p}g_{\mu\nu}$, which is also the relation between the Jordan and the Einstein frames. However, our case is different in that the scalar field is non-minimally coupled to the matter sector, such that the Lagrangian of the model is given as
\beast
S = \int d^4x \sqrt{-g} \; \left(R/2 - {1 \over 2}\; \partial_\mu\phi \partial^\mu\phi - V(\phi) + f(\phi) L_m \right)\;.
\eeast
Here $g$ is the determinant of the metric tensor $g_{\mu\nu}$, $R$ the Ricci scalar constructed from the metric,  $L_m$ is the Lagrangian of the standard matter fields, $V(\phi)$ is the potential of the chameleon-like scalar field $\phi$, and $f(\phi)$ is  some generic function of the scalar field to be described shortly. \\

Assuming a spatially flat Friedmann-Robertson-Walker (FRW) metric, the Friedmann equations of the model are given by
\begin{equation}\label{Friedmann}
3H^2 = \rho_\phi + f(\phi) \rho\;, \qquad 2\dot{H} + 3 H^2 = - p_{\phi} - f(\phi) p\;,
\end{equation}
where $\rho_\phi$ and $p_\phi$ are the energy density and pressure of the scalar field, respectively, given by
\begin{equation}\label{phienergy}
\rho_\phi = {1 \over 2}\; \dot{\phi}^2 + V(\phi), \qquad p_\phi = {1 \over 2}\; \dot{\phi}^2 - V(\phi)\;,
\end{equation}
with an overdot representing the derivative with respect to the cosmic time $t$.
Also, taking the variation of the action with respect to the scalar field $\phi$,  leads to the equation of motion of the scalar field (the generalized Klein-Gordon equation),
\begin{equation}\label{EoM}
\ddot{\phi} + 3H\dot{\phi} + V'(\phi) = f'(\phi) L_m\;,
\end{equation}
where the prime denotes the derivative with respect to the scalar field. The matter Lagrangian in the above equation must be specified in order to close the system of cosmological equation. There are many studies on this topic such as \cite{Carroll:1998zi,Damour:1994zq,Brown:1992kc,Brown:1992bq,Sotiriou:2008it,Saaidi:2013yfa,Aghamohammadi:2013eja,Sheikhahmadi:2018aux,Saaidi:2013pfa}, where the authors explain that there are two definitions for the Lagrangian of a perfect fluid as $L_m^{(1)}=-\rho$ and $L_m^{(2)}=p$, respectively. However, in our case, this degeneracy is broken due to the non-minimal interaction of the scalar field and matter, and therefore we follow the approach proposed in \cite{Saaidi:2013pfa}, where it is shown that only for $L_m^{(2)}=p$ one has a geodesic motion for perfect fluids. Thus, in this work, we are going to adopt this  Lagrangian for the perfect fluid. Because of the interaction between the scalar field and the matter sector, the conservation equations are generalized, and they are given by
\begin{eqnarray}\label{conservation}
{d \over dt} \left[ f(\phi)\rho \right] + 3H f(\phi) (\rho + p) & = & - \dot{f}(\phi) \; p, \nonumber \\
{d \over dt} \rho_\phi + 3H (\rho_\phi + p_\phi) & = & \dot{f}(\phi) p\;.
\end{eqnarray}

\section{Chameleon warm inflation}\label{SecIV}

In this Section, we are going to explain the general features of warm inflation, in which the inflaton is described by a chameleon-like scalar field. In warm inflation a dissipation term appears on the right hand side of the conservation equation, which is usually included by hand. However, in the chameleon model of scalar field one could naturally obtain a dissipation term, as in Eq. \eqref{conservation}, because of the presence of the interaction between the scalar field, and the matter term in the action. In the following, we first review the warm inflationary scenario, and then we  combine the scenario with the chameleon model, where the chameleon scalar field plays the role of the inflaton.

\subsection{Review of warm inflation}\label{SecIVa}

The warm inflationary scenario describes an accelerated expansion phase of the Universe at its earliest times, in which the scalar field is the dominant component. However, its main difference with the cold inflation is that beside a scalar field, there is another component of the cosmological fluid is present, usually taken as radiation. During inflation the scalar field and the matter (radiation) component interact. In such a case, the Friedmann equations are given by
\begin{equation}\label{warmfriedmann}
3H^2 = {1 \over 2} \dot\phi^2 + V(\phi) + \rho_r\;, \qquad 2\dot{H} = -\dot\phi^2 - {4 \over 3} \rho_r\;.
\end{equation}
Due to this interaction, energy is transferred from the scalar field to the radiation fluid,  and this process is described through the following conservation equations,
\begin{eqnarray}\label{warmconservation1}
  \dot{\rho}_r + 3H (\rho_r + p_r) &=& \Gamma \dot\phi^2\;, \\
  \dot{\rho}_\phi + 3H(\rho_\phi + p_\phi) &=& -\Gamma \dot\phi^2\;.\label{warmconservation}
\end{eqnarray}
Here $\Gamma$ is a dissipation coefficient. Eq. \eqref{warmconservation} is usually written in a different way,  known as the equation of motion of the scalar field, as
\begin{equation}\label{warmeom}
\ddot{\phi} + 3H (1+Q) \; \dot{\phi} + V'(\phi) =0\;,
\end{equation}
where the parameter $Q=\Gamma/3H$ is the ratio of the radiation production to expansion rate.

In warm inflation, the slow-roll approximations are still at work, {\it i.e.}, in order to have a quasi-de Sitter expansion, the rate of the Hubble parameter during a Hubble time is assumed to be small, a condition that is imposed via the first slow-roll parameter in Eq. \eqref{srp1}.
The energy density of the scalar field dominates over the radiation energy density, and also the kinetic term of the scalar field is negligible w.r.t its potential, {\it i.e.}  $\rho_\phi \gg  \rho_r$ and $\rho_\phi \simeq V(\phi)$. In addition to the assumptions that we also have in cold inflation, in warm inflationary scenario it is supposed that the radiation production is quasi-stable during inflation, so that $\dot{\rho_r} \ll H\rho_r$ and $\dot{\rho_r} \ll \Gamma \dot\phi^2$. Then, from Eqs. \eqref{warmfriedmann}, \eqref{warmconservation}, and \eqref{warmeom}, it follows that
\begin{eqnarray}
  &&3H^2  \simeq  V(\phi)\;, \label{vfriedmann} \\
 &&\dot\phi  \simeq  - {V'(\phi) \over 3H(1+Q)}\;, \label{dotphi} \\
 &&  \rho_r  =  C_\gamma T^4 = {\Gamma \over 4H} \dot\phi^2\;, \label{radiationtemp}
\end{eqnarray}
where $T$ is the temperature of the fluid, $C_\gamma=\pi^2 g_\star / 30$ is the Stefen-Boltzman constant, and $g_\star$ is the number of degrees of freedom of the radiation field. \\

With the use of the above equations, it follows that the aforementioned slow-roll parameters can be expressed as
\begin{equation}\label{vsrp}
\epsilon_1 = {1 \over 2(1+Q)} {V^{\prime 2}(\phi) \over V^2(\phi)}\;, \qquad \epsilon_2 = {\dot{\epsilon}_1 \over H \epsilon_1}\;.
\end{equation}

It is common to use two other slow-roll parameters, which are stated in terms of the potential as,
\begin{equation}\label{etabeta}
\eta = {1 \over (1+Q)} \; {V'' \over V}, \qquad \beta = {1 \over (1+Q)} \; {V' \Gamma' \over V \Gamma}\;.
\end{equation}

The slow-roll parameter $\epsilon_2$ is related to the above slow-roll parameters through the relation
\begin{equation*}
  \epsilon_2 = -2 \eta + 4 \epsilon_1 + {Q \over (1+Q)} \; \left( \beta - \epsilon_1 \right)\;.
\end{equation*}

Due to the appearance of the term $(1+Q)$ in the dominator of the slow-roll parameters, it follows that the smallness of the parameters are guaranteed for a large range of potentials that could satisfy the slow-roll approximations.

Smallness of the slow-roll parameters ensures us that first the Universe had a quasi-exponential accelerated expansionary phase, and also it did stand in this phase for enough number of e-folds to solve the issues of the hot big bang model. The number of e-folds is given by the following relation,
\begin{equation}\label{efold}
N = \int_{t_\star}^{t_e} H dt = \int_{\phi_\star}^{\phi_e} {H \over \dot\phi} \; d\phi =
- \int_{\phi_\star}^{\phi_e} (1+Q) {V(\phi) \over V'(\phi)} \; d\phi\;,
\end{equation}
where the last equality is obtained by using Eqs. \eqref{dotphi} and \eqref{vfriedmann}. The subscripts $``e"$ and $``\star$"  respectively indicate the quantity at the end of inflation, and at the horizon crossing time, respectively.

\subsection{Warm inflation with a chameleon-like scalar field}\label{warmchameleon}

In the chameleon scalar field model, there is a source term on the right hand side of the conservation equation \eqref{conservation}, which automatically appears in the equation because of the interaction between the scalar field and the matter sector. Since the chameleon scalar field is taken as the inflaton, it follows that our source term in the conservation equation of the chameleon model plays the same role of the term $\Gamma \dot\phi^2$ in Eq. \eqref{warmconservation}. Then, we are going to equate these two terms, {\it i.e.}
\begin{equation}\label{sources}
\Gamma \dot{\phi}^2 = -{1 \over 3} \; \rho_{rs} {\dot{f}(\phi) \over f(\phi)}\;,
\end{equation}
where the second fluid component of the very early Universe is taken as radiation, with equation of state $p_r=\rho_r/3$.
On the other hand, the sign of this interaction also shows up in the Friedmann equation, {\it i.e.}, via
 the presence of the coupling function $f(\phi)$ in Eq. \eqref{Friedmann}. To get back to the original evolution equation, we can define again an effective radiation energy density and pressure as $\rho_{rs}$ and $p_{rs}$. In this case, the evolution equations are reorganized as
\begin{eqnarray}\label{finalconservation}
\dot{\rho}_{rs} + 3H (\rho_{rs} + p_{rs}) &=& -p_{rs} \; {\dot{f}(\phi) \over f(\phi)} =  \Gamma \dot\phi^2\;, \label{finalrconservation} \\
  \dot{\rho}_\phi + 3H(\rho_\phi + p_\phi) &=& p_{rs} \; {\dot{f}(\phi) \over f(\phi)} = -\Gamma \dot\phi^2\;. \label{finalfconservation}
\end{eqnarray}

{Assuming a quasi-stable production of the radiation, so that $\dot\rho_{rs} \ll H\rho_{rs} $ and $\dot\rho_{rs} \ll \Gamma\dot\phi^2$, which implies that $\dot\rho_{rs} \ll p_{rs} \dot{f}(\phi) / f(\phi)$, from} Eq. \eqref{finalrconservation} one can calculate the Hubble parameter as
\begin{equation}\label{hubblef}
  H = -{1 \over 12} \; {\dot{f}(\phi) \over f(\phi)}\;.
\end{equation}
Using Eq. \eqref{vfriedmann}, the potential of the scalar field is derived in terms of $f(\phi)$ and its derivative as
\begin{equation}\label{potf}
  V(\phi) = {1 \over 48} \; \left[ {\dot{f}(\phi) \over f(\phi)} \right]^2\;.
\end{equation}
On the other hand, using the result of QFT in curved space, the dissipation coefficient is taken as its best known expression $\Gamma= \Gamma_0 T^m/ \phi^{m-1}$. Then, by applying Eqs. \eqref{dotphi},  \eqref{radiationtemp} and  \eqref{sources}, the temperature of the radiation fluid is obtained as
\begin{equation}\label{temp}
T^{m-4} = {C_\gamma \over \Gamma_0} \; (1+Q) H \; {f'(\phi) \over f(\phi)} \; {\phi^{m-1} \over V'(\phi)}\;,
\end{equation}
where we have used the mathematical result $\dot{f}(\phi) = f'(\phi) \dot\phi$. \\

To go one step further, we assume the ansatz $\dot{f}(\phi)/f(\phi) = \alpha \phi^n$. With the use of this assumption in Eqs. \eqref{hubblef} and  \eqref{potf}, the Hubble parameter and the potential of the scalar field are obtained as
\begin{equation}\label{hubblepot}
H= -{\alpha \over 12} \; \phi^n, \qquad V(\phi) = {\alpha^2 \over 48} \; \phi^{2n}\;.
\end{equation}
Since the Hubble parameter is positive, $\alpha$ should be a negative constant. \\

On the other hand, with the use of Eqs. \eqref{dotphi},  \eqref{hubblef} and  \eqref{hubblepot} in Eq. \eqref{temp}, one arrives at the following non-linear equation for the temperature of the fluid,
\begin{equation}\label{tempNLequation}
T^m + {\phi^{m+3n-3 \over 2} \over \xi \Gamma_0}\; T^{m-4 \over 2} - {\alpha \over 4\Gamma_0} \phi^{m+n-1}=0\;,
\qquad \xi \equiv \left( - {192 C_\gamma \over n^2 \alpha^3 \Gamma_0} \right)^{1/2}\;.
\end{equation}

Due to the fact that the third term of the above equation is a power-law function of the scalar field, it follows that the temperature $T(\phi)$ should be a power-law function of the scalar field as well, in order to satisfy the equation. Hence, for the temperature we assume the functional form  $T(\phi) = T_0 \phi^q$. By substituting $T$ in Eq. \eqref{temp}, we obtain
\begin{equation*}
T_0^m \phi^{mq} + { T_0^{m-4 \over 2} \over \xi \Gamma_0} \; \phi^{{m+3n-3 \over 2}+{q(m-4) \over 2}} - {\alpha \over 4\Gamma_0} \phi^{m+n-1}=0\;.
\end{equation*}
Then, the power of the scalar field of the three terms should be equal, which gives the following relations:
\begin{enumerate}
  \item[(a)] equality of the first and second terms: $q={m+3n-3 \over m+4}$;
  \item[(b)]  equality of the first and third terms: $q={m+n-1 \over m}$; and
  \item[(c)] equality of the second and third terms: $q={m-n+1 \over m-4}$.
\end{enumerate}

Hence, if the three conditions are satisfied, the proposed function for the temperature might be a proper solution. However, it can be shown that these conditions actually give one constraint that is extracted as follows:
\begin{itemize}
  \item ${\rm a=b}$: $m = {2(n-1) \over n-3}$;
  \item ${\rm a=c}$: $m = {2(n-1) \over n-3}$; and
  \item ${\rm b=c}$: $m = {2(n-1) \over n-3}$.
\end{itemize}
Therefore, for the proposed functional form of the temperature, the following constraints must be satisfied:
\begin{equation}\label{mq}
m={2(n-1) \over n-3}\;, \quad q={n-1 \over 2}\;,
\end{equation}
which show that by determining the power $n$, the power parameters $m$ and $q$ are also determined.

In the warm inflationary scenario, there are both quantum and thermal fluctuations. Thermal fluctuations are dependent on the fluid temperature $T$, and quantum fluctuations depend on the Hubble parameter $H$. A feature of warm inflation is that the fluid temperature is bigger than the Hubble parameter, $T>H$, stating that the thermal fluctuations overcome quantum fluctuations, and become the origin of the Universe's LSS. To ensure that we stay in the warm inflation regime, the condition $T/H > 1$
\begin{equation}\label{TH}
{T(\phi) \over H(\phi)} = -{12 T_0 \over \alpha}\; \phi^{q-n}>1,
\end{equation}
must be satisfied during the cosmological expansion.

The amount of cosmic expansion during  inflation is measured through the number of e-folds $N$ defined as
\begin{equation}\label{efoldf}
N = \int_{t_\star}^{t_e} H \; dt = \int_{\phi_\star}^{\phi_e} {H \over \dot\phi} \; d\phi =-
{1 \over 12} \int_{\phi_\star}^{\phi_e} {f'(\phi)  \over f(\phi)} \; d\phi = -{1 \over 12}\ln\left[ {f(\phi_e) \over f(\phi_\star)} \right].
\end{equation}

Once the expression of the scalar field at the end of inflation is obtained from the relation $\epsilon_1(\phi_e)=1$, Eq. \eqref{efoldf} is used to derive the scalar field at the time of the horizon crossing in terms of the number of e-folds. Then, all the perturbation parameters could be expressed in terms of the number of e-folds. \\

To test the validity of a theoretical model, one has to compare its predictions with the observational data. In this regard, we will obtain some important perturbations parameters, such as the amplitude of scalar perturbations, the scalar spectral index and the tensor-to-scalar ratio,  and they will be compared with the Planck-2018 data. Following \cite{Bastero-Gil:2016qru,Berera:2018tfc}, the amplitude of scalar perturbations is calculated as
\begin{equation}\label{pswarm}
\mathcal{P}_s = \left( { H^2 \over 2\pi \dot\phi } \right)^2 \left(1 + 2n_{BE} + \frac{2\sqrt{3}\pi Q}{\sqrt{3+4\pi Q}}{T \over H}\right) G(Q)\;,
\end{equation}
where $n_{BE}$ is the Bose-Einstein distribution, given by $n_{BE}= \left[ \exp(H/T_{\delta\phi}) - 1 \right]^{-1}$, and $T_{\delta\phi}$ is the temperature of the inflaton fluctuations \cite{Berera:2018tfc}. Here $G(Q)$, giving the growth of the fluctuations, is a function of $Q$ and its presence is due to the coupling of the scalar field and radiation \cite{Bastero-Gil:2016qru,Berera:2018tfc}.
The scalar spectral index and its running are obtained from the amplitude of scalar perturbations, and it is defined as
\begin{equation}\label{nsaswarm}
n_s - 1 = {d\ln(\mathcal{P}_s) \over d\ln(k)}\;, \qquad \alpha_s = {d n_s \over d\ln(k)}\;.
\end{equation}
Tensor perturbations, known as gravitational waves, are measured indirectly through the  tensor-to-scalar ratio parameter $r=\mathcal{P}_t / \mathcal{P}_s$. The amplitude of tensor perturbations is given by \cite{Bastero-Gil:2016qru}
\begin{equation}\label{ptwarm}
\mathcal{P}_t = {2 H^2 \over \pi^2}\;.
\end{equation}
The tensor spectral index is defined as
\begin{equation}\label{ntwarm}
n_t = {d\ln(\mathcal{P}_t) \over d\ln(k)}\;.
\end{equation}

In the relevant literature, the scenario of warm inflation is usually considered in two regimes, known  as the strong and the weak dissipative regimes, where the dissipative ratio is respectively $Q \gg 1$ and $Q \ll 1$. In the subsequent subsections, we are going to consider the discussed model in these two regimes. \\

\subsection{Weak dissipative regime}\label{SecVa}

In the weak dissipative regime, the dissipative ratio is much smaller than unity, {\it i.e.}, $Q\ll1 (\Gamma \ll 3H)$, and thus we have $(1+Q) \simeq 1$. This approximation makes the evolution equation easier to analyse, since the time derivative of the scalar field is expressed as
\begin{equation}\label{dotphiweak}
\dot{\phi} = {n \alpha \over 6} \phi^{n-1}\;.
\end{equation}
Also, by taking  this approximation into account, and following the same process as used previously, the temperature of the fluid is obtained easily as a function of the scalar field:
\begin{equation}\label{tempweak}
T^{m-4}(\phi) = {-12 C_\gamma \over n^2 \alpha \Gamma_0} {1 \over \phi^{n-m-1}}\;.
\end{equation}

In subsection \eqref{warmchameleon}, an ansatz was introduced for the model. Using this definition and applying Eq. \eqref{dotphiweak}, the coupling function $f(\phi)$ is obtained as an exponential function,
\begin{equation}\label{fphiweak}
f(\phi) = f_0 \exp\left( {3 \over n} \; \phi^2 \right)\;,
\end{equation}
where $f_0$ is a constant of integration. \\

The slow-roll parameters, introduced in Eqs. \eqref{vsrp} and  \eqref{etabeta}, are obtained as

\begin{equation}\label{epsilonetaweak}
\epsilon_1(\phi) = {2n^2 \over \phi^2}, \qquad \eta(\phi)= {(2n-1) \over n} \; {1 \over \phi^2}, \qquad
\beta(\phi) = {(6 m-n m-4) \over {n (m-4)}} \epsilon_1(\phi) \;.
\end{equation}

Inflation ends for $\phi_e^2 = 2n^2$, and the scalar field at the horizon crossing is extracted from the number of e-folds. Hence from Eq. \eqref{efoldf} one arrives at the expression,
\begin{equation}\label{efoldweak}
N = \int_{\phi_\star}^{\phi_e} {H \over \dot\phi} \; d\phi = -{1 \over 12} \; \int_{\phi_\star}^{\phi_e} {f'(\phi) \over f(\phi)} d\phi\;,
\end{equation}
which subsequently gives
\begin{equation}\label{phistarweak}
\phi_\star^2 = 2n^2 \left( 1 + {2N \over n} \right)\;.
\end{equation}

In the weak dissipative regimes, the parameter $G(Q) \simeq 1$ at the time of horizon crossing. Then, from Eq. \eqref{pswarm}, the amplitude of scalar perturbations in these regimes is found as
\begin{equation}\label{psweak}
\mathcal{P}_s = 2 \; \left( { H^2 \over 2\pi \dot\phi } \right)^2 {T \over H}\;.
\end{equation}
From Eq. \eqref{nsaswarm} the scalar spectral index and its running in the weak regimes are obtained as
\be\label{nsweak}
 n_s -1= -{17 \over 8} \; \epsilon_1 + {3 \over 2} \; \eta - {1 \over 4} \beta\;,
\ee
where $\eta$ is defined through the slow-roll parameters $\epsilon_2$ as $\eta=V'' / (1+Q)V$, so that $\epsilon_2 =\eta - \epsilon_1$, and the slow-roll parameter $\beta$ is given as $\beta=V' \Gamma' / (1+Q) V \Gamma$. The tensor-to-scalar ratio in this case is also obtained as
\begin{equation}\label{rweak}
r = 8 \epsilon_1 {H \over T}\;.
\end{equation}

To compare the results of the model with observational data, the above perturbation parameters are computed at the horizon crossing. Using Eq. \eqref{phistarweak}, the slow-roll parameters at horizon crossing are obtained as follows:
\begin{eqnarray}\label{srstarweak}
\epsilon_1^\star & = & \left( 1 + {2 N \over n} \right)^{-1}\;, \\
\eta^\star & = & {2n-1 \over n} \epsilon_1^\star\;,  \\
\beta^\star & = & - {nm - 6m +4 \over n(m-4)} \epsilon_1^\star\;.
\end{eqnarray}
Substituting the above slow-roll parameters in Eq. \eqref{nsweak}, the scalar spectral index is given as a function of $n$, $m$ and  e-folding number $N$ as

\begin{equation}\label{nsstarweak}
n_s(n,m,N) = 1 - \left( {17 \over 8} - {3(2n-1) \over 2n} - {nm - 6m +4 \over 4n(m-4)} \right)\left( 1 + {2 N \over n} \right)^{-1}.
\end{equation}
From Eq. \eqref{psweak}, and by using Eq. \eqref{phistarweak}, the model constant parameter $\Gamma_0$ is found in terms of $n$, $m$ and $N$ as

\begin{equation}\label{gammaweak}
\Gamma_0(n,m,N) = -{12 C_\gamma \over \alpha n^2 } \; \left(-{\alpha \over 96 \pi^2 n^2 \mathcal{P}_s} \right)^{m-4}\left[ 2n^2 \left( 1 + {2 N \over n} \right) \right]^{nm -5n+3m-7 \over 2}\;,
\end{equation}
where $\mathcal{P}_s$ is the amplitude of the scalar perturbations at horizon crossing. According to the latest observational data $\mathcal{P}_s = 2.17 \times 10^{-9}$. The tensor-to-scalar ratio for the same time is found from Eq. \eqref{rweak} as

\begin{equation}\label{rstarweak}
r^\star(n,m,N) = -{4 \alpha n^2 \over 3} \; \left(-{\alpha n^2  \over 12 C_\gamma} \right)^{1 \over m-4} \; \Gamma_0^{1 \over m-4}(n,m,N)
                  \; \left[ 2n^2 \left( 1 + {2 N \over n} \right) \right]^{nm -3(n+m)+7 \over 2(m-4)}\;.
\end{equation}

\begin{figure}[ht]
  \centering
  \includegraphics[width=7cm]{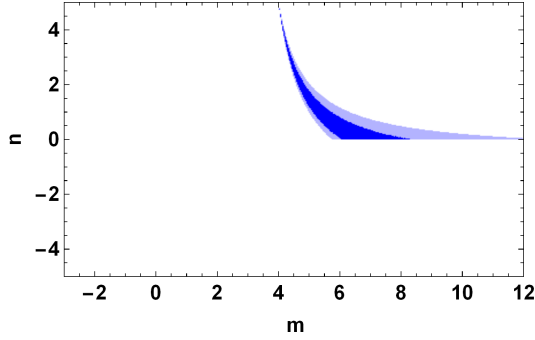}
  \caption{Numerical values of the $(n,m)$ parameters of the chameleon field driven warm inflation model in the weak dissipative regime for which the point $(r-n_s)$ is located in the observational region.  The dark blue color shows the values of $(n,m)$ in the $68\%$ CL range of the Planck-2018 data, while the light blue color shows the parameter values  in the range $95\%$ CL.}\label{nmweak}
\end{figure}

\begin{figure}[ht]
  \centering
  \includegraphics[width=9cm]{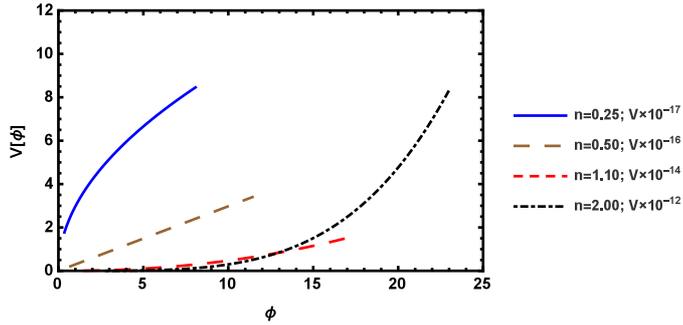}
  \caption{The behavior of the scalar field potential $V(\phi)$ versus $\phi$ of the chameleon field driven warm inflation model in the weak dissipative regime for different values of $n$ during inflation. The figure shows that for $n=2$, the energy scale of inflation is about $10^{15} {\rm GeV}$. Once inflation is approaching to its end, the potential decreases in time. }\label{potweak}
\end{figure}

\begin{figure}[ht]
  \centering
  \includegraphics[width=9cm]{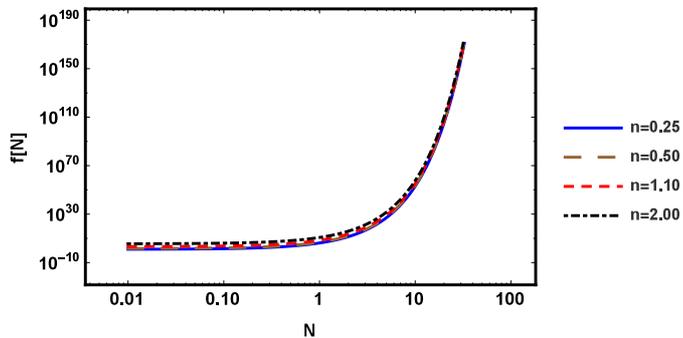}
  \caption{The behavior of the coupling function $f(N)$ versus the number of e-fold for different values of $n$ (in a logarithmic scale). The function has an exponential behavior and when approaching to the end of inflation it rapidly decreases. }\label{fphiNweak}
\end{figure}

Using the $r-n_s$ diagram of Planck-2018, one could plot a $n-m$ diagram as shown in Fig. \ref{nmweak}, where the dark blue color indicates an area of $(n,m)$ where the results for $n_s$ and $r$ stand in $68\%$ CL. The light blue color indicates an area of $(n,m)$ in which the point $(r,n_s)$ of the model stands in $95\%$ CL. \\
Inserting the scalar field at horizon crossing in Eq. \eqref{potf}, the potential of the scalar field could be obtained in terms of $N$ for specific values of $(n,m)$ that have been taken from Fig. \eqref{potweak}. The results in the Figure indicate that the inflation did begin at an energy  scale about $10^{15-16} {\rm{GeV}}$. At the onset of inflation, the scalar field stands on the top, and by passing time it rolls down slowly until the end of inflation. \\
Inserting the scalar field at horizon crossing in Eq. \eqref{fphiweak}, the coupling function could be depicted versus $N$. Getting some point of $(n,m)$ from Fig. \ref{nmweak}, the behavior of the coupling function $f(\phi)$ in the WDR is illustrated in Fig. \ref{fphiNweak}. The curves show the behavior of $\ln\big(f(N)\big)$ versus $\ln(N)$ during the inflationary era; however, the scales on the axes show the actual ranges of $f(N)$ and $N$. The coupling function $f(N)$ has an exponential behavior, and it grows rapidly by enhancement of $N$.

\begin{figure}[ht]
  \centering
  \includegraphics[width=9cm]{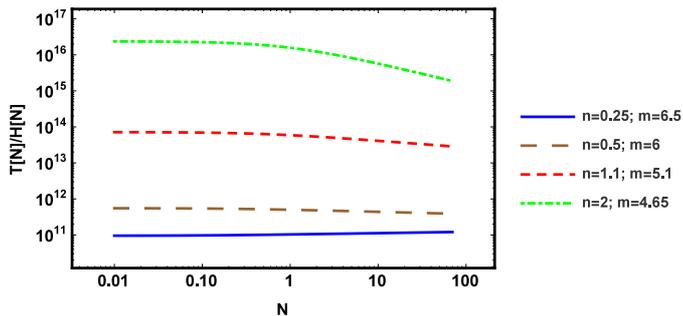}
  \caption{The ratio of the temperature to the Hubble parameter during the inflationary period of the chameleon field driven warm inflation model in the weak dissipative regime versus the number of e-folds for different values of $(n,m)$. As one can see from the plots during inflation the temperature is larger than the Hubble parameter, and the condition $T>H$ is satisfied properly.}\label{thweak}
\end{figure}

 An important feature of the warm inflationary scenario is that the thermal fluctuations overcome the quantum fluctuations, since the fluid temperature is bigger than the Hubble parameter. To have a healthy warm inflation, this condition should be satisfied during the cosmological evolution. Fig. \ref{thweak} describes the behavior of the ratio of the temperature to the Hubble parameter during this era. From the Figure one can see that the condition is satisfied during this phase of cosmological expansion.

\subsection{Strong dissipative regime}\label{SecVb}

In this regime, the dissipative ratio is much larger than unity, so that $(1+Q)\simeq Q$. In this case, the analysis of the model becomes more complicated, and the cosmological dynamics is different from the weak dissipative regime. In this approximation the time derivative of the scalar field is given by
\begin{equation}\label{dotphist}
\dot\phi =- {V'(\phi) \over \Gamma}\;.
\end{equation}
Using Eq. \eqref{sources}, and by considering the definition of $\Gamma$, and Eq. \eqref{dotphist}, the temperature of the radiation fluid is obtained as
\begin{equation}\label{tempstrong}
T^{m+4}(\phi) = -{n^2 \alpha^3 \over 192 \Gamma_0 C_\gamma}\phi^{3n+m-3} = T_0^{m+4} \phi^{3n+m-3}\;.
\end{equation}
Then, with the help of this relation, the time derivative of the scalar field can be found in terms of the scalar field, so that
\begin{equation}\label{dotphistrong}
\dot\phi = -{n \alpha^2 \over 24 \Gamma_0 T_0^m} \phi^{8n-nm+5m-8 \over m+4}\;.
\end{equation}
The dissipative ratio $Q$ is given by
\begin{equation}\label{Qstrong}
Q(\phi) = {\Gamma \over 3H} = -{4 \Gamma_0 T_0^m \over \alpha} \phi^{2(nm-2n-3m+2) \over m+4}.
\end{equation}

From Eqs. \eqref{vsrp} and  \eqref{etabeta}, after substituting Eq. \eqref{Qstrong}, we find out the slow-roll parameters as

\begin{eqnarray}
  \epsilon_1(\phi) &=& -{\alpha n^2 \over 2 \Gamma_0 T_0^m} \phi^{-2(nm-2n-2m+6) \over m+4}\;, \\
  \eta(\phi) &=&  { (2n-1) \over n} \; \epsilon_1(\phi)\;,\\
  \beta(\phi) &=&  {m(3n-6)+4 \over n (m+4)} \; \epsilon_1(\phi)\;.
\end{eqnarray}
Then, the scalar field at the end of inflation is obtained as
\begin{equation*}
  \phi_e^{2(nm-2n-2m+6) \over m+4} = -{\alpha n^2 \over 2 \Gamma_0 T_0^m}.
\end{equation*}
After obtaining $\dot\phi(\phi)$ and $Q(\phi)$, the coupling function $f(\phi)$ can also be determined through the definition introduced previously. It turns out that $f(\phi)$ is again an exponential function of the scalar field, and it is given by
\begin{equation}\label{fphistrong}
f(\phi) = f_0 \exp\left( -{12 \Gamma_0 T_0^m \over n \alpha} \; {m+4 \over nm - 2n - 2m + 6} \; \phi^{2(nm - 2n - 2m + 6) \over m+4} \right),
\end{equation}
where $f_0$ is an arbitrary constant of integration. Using this result in the relation of $N$, after integration, the scalar field at the time of horizon crossing is obtained in terms of $N$ as
\begin{equation}\label{phistarstrong}
\phi^{2(nm-2n-2m+6) \over m+4}_\star = -{\alpha n^2 \over 2 \Gamma_0 T_0^m} \; \left( 1 + {2(nm-2n-2m+6) \over n (m+4)}\; N \right)\;.
\end{equation}

In the strong dissipative regime, the parameter $G(Q)$, depending on the different values of the parameter $m$, can be approximated as \cite{Bastero-Gil:2016qru,Berera:2018tfc}:
\begin{eqnarray*}
m=1  & \longrightarrow & G(Q)\simeq 1+ 0.127 Q^{4.330}+ 4.981Q^{1.946},,\\
m=3  & \longrightarrow & G(Q)\simeq 1+ 0.0185Q^{2.315}+ 0.335 Q^{1.364},\\
m=-1 & \longrightarrow & G(Q)\simeq \frac{1+ 0.4 Q^{0.77}}{(1+0.15Q^{1.09})^2}.
\end{eqnarray*}
Therefore, in a more convenient way, we can write down the function $G(Q)=a_mQ^{b_m}$, where
 \begin{eqnarray*}
m=1   & \longrightarrow &  a_m=0.127, \quad  b_m={4.330},\\
m=3   & \longrightarrow &  a_m=0.0185, \quad  b_m={2.315},\\
m=-1  & \longrightarrow &  a_m =17.78, \quad  b_m={-1.41}\;.\\
\end{eqnarray*}
{For the case $m=-1$, the power of the term $Q$ is negative stating that the term $G(Q)$ tends to zero which in turn lead the amplitude of the scalar perturbation to zero. So, this case is not considered here.} \\
Hence the amplitude of the scalar perturbations is obtained as
\begin{equation}\label{psstrong}
  \mathcal{P}_s = \left( { H^2 \over 2\pi \dot\phi } \right)^2 \; \sqrt{3\pi Q} \; {T \over H} \times a_mQ^{b_m}\;.
\end{equation}
By taking the time derivative of this equation according to Eq. \eqref{nsaswarm} leads to the scalar spectral index and its running, given by
\begin{eqnarray}\label{nsasstrong}
  n_s -1 &=& -2 q_1 \epsilon_1  - q_2 \beta + {3 \over 2} \eta,
\end{eqnarray}
where $q_1=\frac{9}{8}- {b_m \over 2}$ and $q_2=b_m+\frac{7}{4}$. We notice that the function $G(Q)$ for $m=3$ has an acceptable behaviour for the value  $m=-1$, or even for $m=1$.
The amplitude of tensor perturbations in strong dissipative regimes is given by \cite{Bastero-Gil:2016qru,Berera:2018tfc}
\begin{equation}\label{ptstrong}
\mathcal{P}_t = {2 H^2 \over \pi^2} = {\alpha^2 \over 72\pi^2} \phi^{2n}\;.
\end{equation}
Then, the tensor-to-scalar ratio is obtained as follows:
\begin{equation}\label{rstrong}
r =\frac{{24{{(\sqrt 3 )}^{{b_m}}}}}{{\sqrt {3\pi } \,{a_m}}}{(\frac{{4{C_\gamma }}}{9})^{\frac{1}{4}}}\frac{{{{V'}^{3/2}}}}{{{V^{({q_1} - 1)}}{\Gamma ^{{q_2}}}}}\;.
\end{equation}

Using Eq. \eqref{phistarstrong}, the above perturbation parameters can be obtained at the horizon crossing time. The slow-roll parameters in the SDR at this cosmological instance are given by
\begin{eqnarray}\label{srstarstrong}
\epsilon_1^\star & = & \left( 1 + {2 (nm-2n-2m+6) N \over n(m+4)} \right)^{-1}\;, \\
\eta^\star & = & {2n-1 \over n} \; \epsilon_1^\star \;, \\
\beta^\star & = & {3nm-6m+4 \over n(m+4)} \; \epsilon_1^\star\;.
\end{eqnarray}
Hence we obtain the following scalar spectral index,
\begin{eqnarray}\label{nsstarstrong}
n_s(n,\alpha,N) & = & 1 - \left( 2q_1\;+\frac{3nm-6m+4}{n(m+4)} \; q_2 - \frac{3(2n-1)}{2n} \right) \epsilon_1^\star\;.
\end{eqnarray}

We notice that this is a function of $n$, $m$ and  $N$. The constant parameter $\Gamma_0$ can be determined by using the observational data for the amplitude of the scalar perturbations as
\begin{eqnarray}\label{gammastrong}
{\Gamma _0}(n,\alpha,N) = \frac{B}{{{\cal P}_s^ \star }}{\left( { - \frac{{{n^2}\alpha }}{2}\;{{\left[- {\frac{{192{C_\gamma }}}{{ {n^2}{\alpha ^3}}}} \right]}^{\frac{m}{{m + 4}}}}\frac{1}{{\epsilon_1^ \star }}} \right)^{\frac{{(m + 4)\sigma }}{{2S(nm - 2n - 2m + 6)}}}} \;,
\end{eqnarray}
where
\begin{equation*}
B(n,\alpha,N) = \frac{1}{{{{(2n)}^{3/2}}}}{(\frac{9}{{4{C_\gamma }}})^{\frac{1}{4}}}\frac{{\sqrt {3\pi } {\mkern 1mu} {a_m}}}{{36{{(\sqrt 3 )}^{{b_m}}}{\pi ^2}}}{(\frac{{{\alpha ^2}}}{{48}})^{{q_1} - \frac{3}{2}}}{\left[- {\frac{{192{C_\gamma }}}{{ {n^2}{\alpha ^3}}}} \right]^{\frac{{ - m{q_2}}}{{m + 4}}}}\;,
\end{equation*}
\begin{equation*}
S(n,\alpha,N)= \frac{{4{q_2}}}{{m + 4}} - \frac{{4\sigma }}{{(nm - 2n - 2m + 6)}}\;,
\end{equation*}
and
\begin{equation*}
\sigma(n,\alpha,N)  =  - \frac{{3(2n - 1)}}{2} + 2n{q_1} + \frac{{(3nm - 6m + 4)}}{{m + 4}}{q_2}\;,
\end{equation*}
respectively. Using the above relations, the tensor-to-scalar ratio at the horizon crossing is obtained as

\begin{eqnarray}\label{rstarstrong}
{r^ \star }(n,\alpha,N) = \frac{{{\alpha ^2}}}{{72{\pi ^2}}}\frac{{{B^{ - 1}}}}{{{\cal P}_s^ \star }}{\left( { - \frac{{{n^2}\alpha }}{2}\;{{\left[ {\frac{{192{C_\gamma }}}{{ - {n^2}{\alpha ^3}}}} \right]}^{\frac{m}{{m + 4}}}}\frac{1}{{\epsilon_1^ \star }}} \right)^{\frac{{(m + 4)(2n - \sigma )}}{{2(nm - 2n - 2m + 6)}}}}\Gamma _0^{\tilde S},
\end{eqnarray}
where we have defined
\begin{equation*}
\tilde S(n,\alpha,N) =  - \frac{{4{q_2}}}{{m + 4}} - \frac{{2(2n - \sigma )}}{{(nm - 2n - 2m + 6)}}\;.
\end{equation*}

\begin{figure}[ht]
  \centering
  \includegraphics[width=7cm]{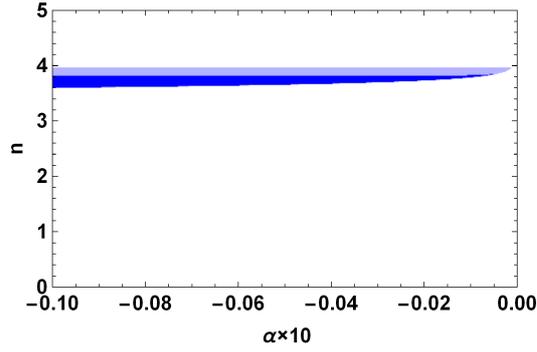}
  \caption{The ranges of the parameters $(n,\alpha)$ of the chameleon field driven warm inflation model in the strong dissipative regime that are consistent with $n_s$ and $r$  located in the observational area. The light blue color indicates the areas with $95\%$ CL, while the dark blue color indicates the areas with $68\%$ CL.}\label{nmstrong}
\end{figure}

\begin{figure}[ht]
  \centering
  \includegraphics[width=9cm]{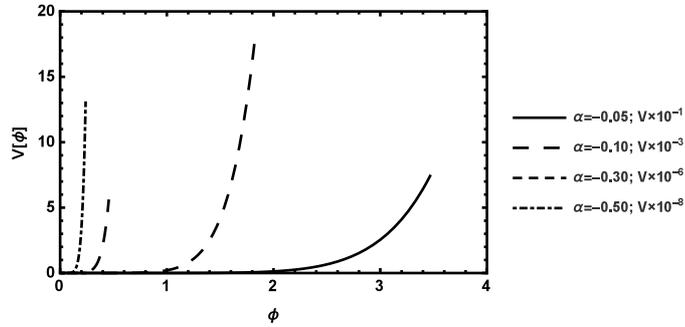}
  \caption{The behavior of the scalar field potential $V(\phi)$ versus the inflation scalar field in the chameleon field driven warm inflation model in the strong dissipative regime for different values of $(n,\alpha)$, obtained from Fig. \ref{nmstrong}. The figure shows that the inflation could start from energy scales of the order of $10^{15} {\rm GeV}$, with the scalar field potential decreasing during the cosmological evolution.}\label{potstrong}
\end{figure}

\begin{figure}[ht]
  \centering
  \includegraphics[width=9.7cm]{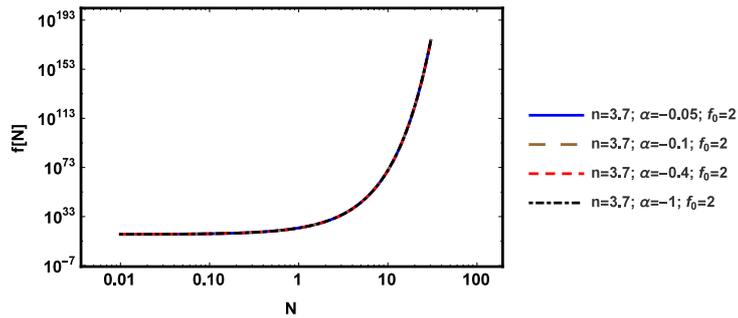}
  \caption{The behavior of coupling function $f(N)$ versus the number of e-folds in the chameleon field driven warm inflation model in the strong dissipative regime for different values of $(n,\alpha)$, which are selected from Fig. \ref{nmstrong}. The function $f(N)$ has an exponential behavior, and during inflation it decreases.}\label{fphiNstrong}
\end{figure}

Similarly as in the previous Section, we use the $r-n_s$ diagram of Planck-2018 to find out the proper region of $(n,\alpha)$, which leads for the model prediction to be agreement with observational data. Fig. \ref{nmstrong} shows this area, depicted with dark and light blue colors, respectively, displaying the values of $(n,\alpha)$ compatible with $68\%$ and $95\%$ CL, respectively. \\
With the use of the value of the scalar field at horizon crossing, {\it i.e.} substituting Eq. \eqref{phistarstrong} in Eq. \eqref{potf}, and by choosing the proper values of $(n,\alpha)$ from Fig. \ref{nmstrong}, the scalar field potential is expressed in terms of the number of e-folds. Fig. \ref{potstrong} depicts the behavior of the scalar field potential during inflation, indicating that the inflation energy scale could be about $10^{15-16} {\rm GeV}$.
By using the value of the scalar field at horizon crossing, {\it i.e.} Eq. \eqref{phistarstrong} in Eq. \eqref{fphistrong}, the coupling function $f(\phi)$ can again be expressed in terms of the number of e-folds. Fig. \ref{fphiNstrong} portrays the behavior of the coupling function during inflation for different values of $(n,\alpha)$ that have been selected from Fig. \ref{nmstrong}.  The curves depict the behavior of $\ln\left[f(N)\right]$ versus $\ln(N)$ during the inflationary times, however, the scales show the true values of $f(N)$ and $N$. The coupling function $f(N)$ has an exponential behavior and it grows rapidly by the enhancement of $N$.

\begin{figure}[ht]
  \centering
  \includegraphics[width=9.7cm]{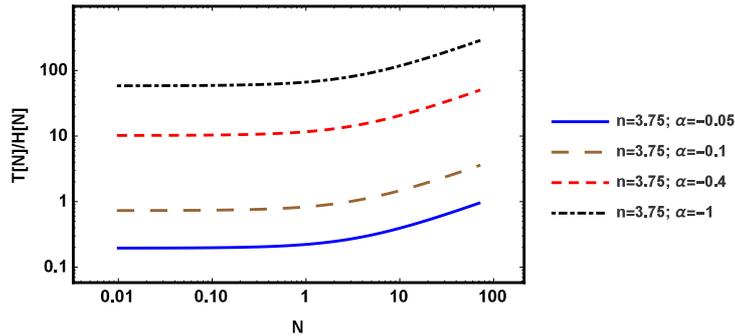}
  \caption{The ratio of the temperature to the Hubble parameter during the inflationary evolution of the chameleon field driven warm inflation model in the strong dissipative regime versus the number of e-folds for different values of $(n,\alpha)$. During inflation the temperature is bigger than the Hubble parameter, and their ratio increases when approaching the end of inflation.}\label{thstrong}
\end{figure}

As the final step, the condition $T/H > 1$ is considered by plotting the ratio of the temperature to the Hubble parameter versus the number of e-folds during inflation, as shown in Fig. \ref{thstrong}.  From the Figure it immediately follows that the condition is satisfied during the cosmological evolution.

\section{Conclusion}\label{conclusion}

In the present work, we have studied the behavior of chameleon-like scalar fields, non-minimally interacting with matter, as potential candidates for driving  warm inflationary evolution. Since the scalar field interacts with the matter sector, in these models one has a generalized energy conservation law, where there is a source  for the scalar field, depending on the scalar field - matter coupling function, and on the matter Lagrangian. Besides, due to this interaction, the mass of the scalar field strongly depends on the environment energy density, so that on cosmological length scales, the scalar field mass is small, while on short scales, like, for example,  around the Earth, the scalar field mass grows. This interesting feature of the chameleon scalar field model may be of major interest  for cosmology.\\

The scenario of warm inflation is a different story compared to the cold one. Although this model has almost the same basic assumptions, here the scalar field interacts with the matter sector during inflation, and energy transfer from the scalar field to the matter part does occur during the accelerated expansion phase. Due to this feature, the chameleon scalar field model seems to be a suitable model for describing warm inflation, since it naturally predicts such an interaction. By using quantum field theory and particle physics, it can be shown that the dissipation coefficient, in general, depends on both scalar field and temperature. Applying this result, and imposing an appropriate ansatz, the Hubble parameter, the scalar field potential, the coupling function, and the temperature can be obtained in terms of the scalar field. Then, the chameleon field driven warm inflationary scenario was considered in detail in two approximations, corresponding to the weak and strong dissipative regimes, respectively. The main perturbation parameters, such as the amplitude of scalar perturbations, the scalar spectral index, and tensor-to-scalar ratio were derived for each regime. After computing the scalar spectral index and tensor-to-scalar ratio, and by using the $r-n_s$ diagram of Planck-2018, we plotted the area of the model parameters $(n,m)$ and $(n,\alpha)$ that places the model exactly in the $68\%$ and $95\%$ CL areas, which are depicted in Figs. \ref{nmweak} and \ref{nmstrong}, respectively, corresponding to the weak and strong dissipative regimes.

Using the results on the model parameters, the scalar field potentials were obtained for both regimes. Our results show that the inflation energy scale could be around $10^{15} {\rm GeV}$, and the scalar field stands on the top of its potential, and rolls down slowly during the cosmological evolution, when and approaching to the end of inflation. Also, the results of the investigation of the coupling function $f(\phi)$ shows that it has an exponential behavior during inflation in both regimes. \\

Another difference between warm inflation and cold inflation is about the type of perturbations. In cold inflation, the quantum perturbations are produced during inflation, and these perturbation are the seeds for LSS of the Universe. On the other hand, in warm inflation, there are both quantum and thermal perturbations, and it is the thermal perturbations that dominate, and are the basic seeds for the formation of the LSS of the Universe. This domination is expressed by the condition $T>H$, otherwise, the quantum fluctuations are overwhelmed and we do not have warm inflation anymore. So, this condition as an important feature of warm inflation was studied as well.  Our investigations show that during inflation the temperature is larger than the Hubble parameter, and  this result has been illustrated in Figs. \ref{thweak} and \ref{thstrong}, respectively.

 The initial conditions of the inflation, as well as the matter content at the moment of the birth of the Universe are not fully known. Similarly, there are many theoretical and observational uncertainties  in the physics of the decay and interactions of the scalar fields. Probably a full understanding of these processes may need physics beyond the Standard Model of particle physics to account for the radiation and other forms of matter creation process. And,  to constrain the theoretical models, and find out what truly happened in the very early
Universe. certainly more accurate observational data are needed. In the present investigation we have proposed some basic tools that may open the possibility of an in-depth comparison of the predictions of the theoretical models of inflation with the cosmological data.

\section*{Acknowledgment}\label{secconAck}
HS would like to thank G. Ellis, A. Weltman, and UCT for arranging his short visit, and the good discussions about fluctuations and perturbations for both large and local scales.
He is thankful to H. Firouzjahi for his good and constructive comments on the initial version of this work. He is also grateful IPM for their hospitality and partial support. He  also wants to thank  ICTP, during Summer School 2018, for the enlightening ideas about perturbation theory. His thanks to his wife E. Avirdi for her patience during our stay in South Africa.
 AM would like to thank the Ministry of Science, Research and Technology of Iran for financial support during his visit to the University of Bologna. He also thanks Prof. R. Casadio and the department of Physics and Astronomy of the University of Bologna for their kind hospitality.
 Christian Corda has been supported financially by the Research Institute for Astronomy and Astrophysics of Maragha (RIAAM). AA acknowledges that this work is based on the research supported in part by the National Research Foundation (NRF) of South Africa (grant numbers 109257 and 112131).






\begin{thebibliography}{99}
\bibitem{Guth:1980zm}
A.~H. Guth, {\it {The Inflationary Universe: A Possible Solution to the Horizon
  and Flatness Problems}},  {\em Phys. Rev.} {\bf D23} (1981) 347--356.

\bibitem{Linde:1981mu}
A.~D. Linde, {\it {A New Inflationary Universe Scenario: A Possible Solution of
  the Horizon, Flatness, Homogeneity, Isotropy and Primordial Monopole
  Problems}},  {\em Phys.Lett.} {\bf B108} (1982) 389--393.

\bibitem{Albrecht:1982wi}
A.~Albrecht and P.~J. Steinhardt, {\it {Cosmology for Grand Unified Theories
  with Radiatively Induced Symmetry Breaking}},  {\em Phys.Rev.Lett.} {\bf 48}
  (1982) 1220--1223.

\bibitem{Linde:1983gd}
A.~D. Linde, {\it {Chaotic Inflation}},  {\em Phys. Lett.} {\bf B129} (1983)
  177--181.

\bibitem{Linde:2007fr}
A.~D. Linde, {\it {Inflationary Cosmology}},  {\em Lect. Notes Phys.} {\bf 738}
  (2008) 1--54, [\href{http://xxx.lanl.gov/abs/0705.0164}{{\tt
  arXiv:0705.0164}}].

\bibitem{Kazanas:1980tx}
  D.~Kazanas,
 \emph{Dynamics of the Universe and Spontaneous Symmetry Breaking},
 { \em Astrophys.\ J.\ }  {\bf 241},  (1980) L59.

\bibitem{Martin:2003bt}
J.~Martin, {\it {Inflation and precision cosmology}},  {\em Braz. J. Phys.}
  {\bf 34} (2004) 1307--1321,
  [\href{http://xxx.lanl.gov/abs/astro-ph/0312492}{{\tt astro-ph/0312492}}].

\bibitem{Martin:2004um}
J.~Martin, {\it {Inflationary cosmological perturbations of quantum- mechanical
  origin}},  {\em Lect. Notes Phys.} {\bf 669} (2005) 199--244,
  [\href{http://xxx.lanl.gov/abs/hep-th/0406011}{{\tt hep-th/0406011}}].

\bibitem{Martin:2007bw}
J.~Martin, {\it {Inflationary perturbations: The cosmological Schwinger
  effect}},  {\em Lect. Notes Phys.} {\bf 738} (2008) 193--241,
  [\href{http://xxx.lanl.gov/abs/0704.3540}{{\tt arXiv:0704.3540}}].
\bibitem{Starobinsky:1979ty}
A.~A. Starobinsky, {\it {Relict Gravitation Radiation Spectrum and Initial
  State of the Universe. (In Russian)}},  {\em JETP Lett.} {\bf 30} (1979)
  682--685.

\bibitem{Mukhanov:1981xt}
V.~F. Mukhanov and G.~Chibisov, {\it {Quantum Fluctuation and Nonsingular
  Universe. (In Russian)}},  {\em JETP Lett.} {\bf 33} (1981) 532--535.

\bibitem{Hawking:1982cz}
S.~Hawking, {\it {The Development of Irregularities in a Single Bubble
  Inflationary Universe}},  {\em Phys. Lett.} {\bf B115} (1982) 295. Revised
  version.

\bibitem{Starobinsky:1982ee}
A.~A. Starobinsky, {\it {Dynamics of Phase Transition in the New Inflationary
  Universe Scenario and Generation of Perturbations}},  {\em Phys. Lett.} {\bf
  B117} (1982) 175--178.

\bibitem{Guth:1982ec}
A.~H. Guth and S.~Y. Pi, {\it {Fluctuations in the New Inflationary Universe}},
   {\em Phys. Rev. Lett.} {\bf 49} (1982) 1110--1113.

\bibitem{Bardeen:1983qw}
J.~M. Bardeen, P.~J. Steinhardt, and M.~S. Turner, {\it {Spontaneous Creation
  of Almost Scale - Free Density Perturbations in an Inflationary Universe}},
  {\em Phys. Rev.} {\bf D28} (1983) 679.

\bibitem{Martin:2013tda}
  J.~Martin, C.~Ringeval and V.~Vennin,
{\it Encyclop{\ae}dia Inflationaris},
  {\em Phys.\ Dark Univ.\ }  {\bf 5-6}, 75 (2014)
  doi:10.1016/j.dark.2014.01.003
  [arXiv:1303.3787 [astro-ph.CO]].


\bibitem{Stewart:1993bc}
E.~D. Stewart and D.~H. Lyth, {\it {A More accurate analytic calculation of the
  spectrum of cosmological perturbations produced during inflation}},  {\em
  Phys. Lett.} {\bf B302} (1993) 171--175,
  [\href{http://xxx.lanl.gov/abs/gr-qc/9302019}{{\tt gr-qc/9302019}}].



\bibitem{Mukhanov:1990me}
V.~F. Mukhanov, H.~A. Feldman, and R.~H. Brandenberger, {\it {Theory of
  cosmological perturbations. Part 1. Classical perturbations. Part 2. Quantum
  theory of perturbations. Part 3. Extensions}},  {\em Phys. Rept.} {\bf 215}
  (1992) 203--333.

\bibitem{Liddle:2000cg}
A.~R.~Liddle and D.~H.~Lyth,
{\it{Cosmological inflation and large scale structure}},
  Cambridge, UK: Univ. Pr. (2000) 400 p


\bibitem{Liddle:1994dx}
A.~R. Liddle, P.~Parsons, and J.~D. Barrow, {\it {Formalizing the slow roll
  approximation in inflation}},  {\em Phys. Rev.} {\bf D50} (1994) 7222--7232,
  [\href{http://xxx.lanl.gov/abs/astro-ph/9408015}{{\tt astro-ph/9408015}}].



\bibitem{Bennett:2012fp}
C.~Bennett, D.~Larson, J.~Weiland, N.~Jarosik, G.~Hinshaw, et~al., {\it
  {Nine-Year Wilkinson Microwave Anisotropy Probe (WMAP) Observations: Final
  Maps and Results}},  \href{http://xxx.lanl.gov/abs/1212.5225}{{\tt
  arXiv:1212.5225}}.

\bibitem{Hinshaw:2012fq}
G.~Hinshaw, D.~Larson, E.~Komatsu, D.~Spergel, C.~Bennett, et~al., {\it
  {Nine-Year Wilkinson Microwave Anisotropy Probe (WMAP) Observations:
  Cosmological Parameter Results}},
  \href{http://xxx.lanl.gov/abs/1212.5226}{{\tt arXiv:1212.5226}}.

\bibitem{Wang:1999vf}
L.-M. Wang and M.~Kamionkowski, {\it {The Cosmic microwave background
  bispectrum and inflation}},  {\em Phys.Rev.} {\bf D61} (2000) 063504,
  [\href{http://xxx.lanl.gov/abs/astro-ph/9907431}{{\tt astro-ph/9907431}}].



  \bibitem{Gangui:1993tt}
A.~Gangui, F.~Lucchin, S.~Matarrese, and S.~Mollerach, {\it {The Three point
  correlation function of the cosmic microwave background in inflationary
  models}},  {\em Astrophys.J.} {\bf 430} (1994) 447--457,
  [\href{http://xxx.lanl.gov/abs/astro-ph/9312033}{{\tt astro-ph/9312033}}].

\bibitem{Gangui:1994yr}
A.~Gangui, {\it {NonGaussian effects in the cosmic microwave background from
  inflation}},  {\em Phys.Rev.} {\bf D50} (1994) 3684--3691,
  [\href{http://xxx.lanl.gov/abs/astro-ph/9406014}{{\tt astro-ph/9406014}}].




\bibitem{Kiefer:1998qe}
C.~Kiefer, D.~Polarski, and A.~A. Starobinsky, {\it {Quantum to classical
  transition for fluctuations in the early universe}},  {\em Int.J.Mod.Phys.}
  {\bf D7} (1998) 455--462, [\href{http://xxx.lanl.gov/abs/gr-qc/9802003}{{\tt
  gr-qc/9802003}}].

\bibitem{Kiefer:2008ku}
C.~Kiefer and D.~Polarski, {\it {Why do cosmological perturbations look
  classical to us?}},  {\em Adv.Sci.Lett.} {\bf 2} (2009) 164--173,
  [\href{http://xxx.lanl.gov/abs/0810.0087}{{\tt arXiv:0810.0087}}].

\bibitem{Sudarsky:2009za}
D.~Sudarsky, {\it {Shortcomings in the Understanding of Why Cosmological
  Perturbations Look Classical}},  {\em Int.J.Mod.Phys.} {\bf D20} (2011)
  509--552, [\href{http://xxx.lanl.gov/abs/0906.0315}{{\tt arXiv:0906.0315}}].

\bibitem{Martin:2012pea}
J.~Martin, V.~Vennin, and P.~Peter, {\it {Cosmological Inflation and the
  Quantum Measurement Problem}},  {\em Phys.Rev.} {\bf D86} (2012) 103524,
  [\href{http://xxx.lanl.gov/abs/1207.2086}{{\tt arXiv:1207.2086}}].

\bibitem{Martin:2012ua}
J.~Martin, {\it {The Quantum State of Inflationary Perturbations}},  {\em
  J.Phys.Conf.Ser.} {\bf 405} (2012) 012004,
  [\href{http://xxx.lanl.gov/abs/1209.3092}{{\tt arXiv:1209.3092}}].

\bibitem{Alexander:2000xv}
S.~Alexander, R.~H. Brandenberger, and D.~Easson, {\it {Brane gases in the
  early universe}},  {\em Phys.Rev.} {\bf D62} (2000) 103509,
  [\href{http://xxx.lanl.gov/abs/hep-th/0005212}{{\tt hep-th/0005212}}].

\bibitem{Steinhardt:2001st}
P.~J. Steinhardt and N.~Turok, {\it {Cosmic evolution in a cyclic universe}},
  {\em Phys.Rev.} {\bf D65} (2002) 126003,
  [\href{http://xxx.lanl.gov/abs/hep-th/0111098}{{\tt hep-th/0111098}}].

\bibitem{Khoury:2001bz}
J.~Khoury, B.~A. Ovrut, N.~Seiberg, P.~J. Steinhardt, and N.~Turok, {\it {From
  big crunch to big bang}},  {\em Phys.Rev.} {\bf D65} (2002) 086007,
  [\href{http://xxx.lanl.gov/abs/hep-th/0108187}{{\tt hep-th/0108187}}].

\bibitem{Khoury:2001wf}
J.~Khoury, B.~A. Ovrut, P.~J. Steinhardt, and N.~Turok, {\it {The Ekpyrotic
  universe: Colliding branes and the origin of the hot big bang}},  {\em
  Phys.Rev.} {\bf D64} (2001) 123522,
  [\href{http://xxx.lanl.gov/abs/hep-th/0103239}{{\tt hep-th/0103239}}].

\bibitem{Martin:2001ue}
J.~Martin, P.~Peter, N.~Pinto~Neto, and D.~J. Schwarz, {\it {Passing through
  the bounce in the ekpyrotic models}},  {\em Phys.Rev.} {\bf D65} (2002)
  123513, [\href{http://xxx.lanl.gov/abs/hep-th/0112128}{{\tt
  hep-th/0112128}}].

\bibitem{Steinhardt:2002ih}
P.~Steinhardt and N.~Turok, {\it {A cyclic model of the universe}},  {\em
  Science} {\bf 296} (2002) 1436--1439.

\bibitem{Finelli:2001sr}
F.~Finelli and R.~Brandenberger, {\it {On the generation of a scale invariant
  spectrum of adiabatic fluctuations in cosmological models with a contracting
  phase}},  {\em Phys.Rev.} {\bf D65} (2002) 103522,
  [\href{http://xxx.lanl.gov/abs/hep-th/0112249}{{\tt hep-th/0112249}}].

\bibitem{Brandenberger:2001kj}
R.~Brandenberger, D.~A. Easson, and D.~Kimberly, {\it {Loitering phase in brane
  gas cosmology}},  {\em Nucl.Phys.} {\bf B623} (2002) 421--436,
  [\href{http://xxx.lanl.gov/abs/hep-th/0109165}{{\tt hep-th/0109165}}].

\bibitem{Kallosh:2001ai}
R.~Kallosh, L.~Kofman, and A.~D. Linde, {\it {Pyrotechnic universe}},  {\em
  Phys.Rev.} {\bf D64} (2001) 123523,
  [\href{http://xxx.lanl.gov/abs/hep-th/0104073}{{\tt hep-th/0104073}}].

\bibitem{Martin:2002ar}
J.~Martin, P.~Peter, N.~Pinto-Neto, and D.~J. Schwarz, {\it {Comment on'Density
  perturbations in the ekpyrotic scenario'}},  {\em Phys.Rev.} {\bf D67} (2003)
  028301, [\href{http://xxx.lanl.gov/abs/hep-th/0204222}{{\tt
  hep-th/0204222}}].

\bibitem{Peter:2002cn}
P.~Peter and N.~Pinto-Neto, {\it {Primordial perturbations in a non singular
  bouncing universe model}},  {\em Phys.Rev.} {\bf D66} (2002) 063509,
  [\href{http://xxx.lanl.gov/abs/hep-th/0203013}{{\tt hep-th/0203013}}].

\bibitem{Tsujikawa:2002qc}
S.~Tsujikawa, R.~Brandenberger, and F.~Finelli, {\it {On the construction of
  nonsingular pre - big bang and ekpyrotic cosmologies and the resulting
  density perturbations}},  {\em Phys.Rev.} {\bf D66} (2002) 083513,
  [\href{http://xxx.lanl.gov/abs/hep-th/0207228}{{\tt hep-th/0207228}}].

\bibitem{Kofman:2002cj}
L.~Kofman, A.~D. Linde, and V.~F. Mukhanov, {\it {Inflationary theory and
  alternative cosmology}},  {\em JHEP} {\bf 0210} (2002) 057,
  [\href{http://xxx.lanl.gov/abs/hep-th/0206088}{{\tt hep-th/0206088}}].

\bibitem{Khoury:2003rt}
J.~Khoury, P.~J. Steinhardt, and N.~Turok, {\it {Designing cyclic universe
  models}},  {\em Phys.Rev.Lett.} {\bf 92} (2004) 031302,
  [\href{http://xxx.lanl.gov/abs/hep-th/0307132}{{\tt hep-th/0307132}}].

\bibitem{Martin:2003bp}
J.~Martin and P.~Peter, {\it {On the causality argument in bouncing
  cosmologies}},  {\em Phys.Rev.Lett.} {\bf 92} (2004) 061301,
  [\href{http://xxx.lanl.gov/abs/astro-ph/0312488}{{\tt astro-ph/0312488}}].

\bibitem{Martin:2003sf}
J.~Martin and P.~Peter, {\it {Parametric amplification of metric fluctuations
  through a bouncing phase}},  {\em Phys.Rev.} {\bf D68} (2003) 103517,
  [\href{http://xxx.lanl.gov/abs/hep-th/0307077}{{\tt hep-th/0307077}}].

\bibitem{Martin:2004pm}
J.~Martin and P.~Peter, {\it {On the properties of the transition matrix in
  bouncing cosmologies}},  {\em Phys.Rev.} {\bf D69} (2004) 107301,
  [\href{http://xxx.lanl.gov/abs/hep-th/0403173}{{\tt hep-th/0403173}}].

\bibitem{Nayeri:2005ck}
A.~Nayeri, R.~H. Brandenberger, and C.~Vafa, {\it {Producing a scale-invariant
  spectrum of perturbations in a Hagedorn phase of string cosmology}},  {\em
  Phys.Rev.Lett.} {\bf 97} (2006) 021302,
  [\href{http://xxx.lanl.gov/abs/hep-th/0511140}{{\tt hep-th/0511140}}].

\bibitem{Peter:2006hx}
P.~Peter, E.~J. Pinho, and N.~Pinto-Neto, {\it {A Non inflationary model with
  scale invariant cosmological perturbations}},  {\em Phys.Rev.} {\bf D75}
  (2007) 023516, [\href{http://xxx.lanl.gov/abs/hep-th/0610205}{{\tt
  hep-th/0610205}}].

\bibitem{Finelli:2007tr}
F.~Finelli, P.~Peter, and N.~Pinto-Neto, {\it {Spectra of primordial
  fluctuations in two-perfect-fluid regular bounces}},  {\em Phys.Rev.} {\bf
  D77} (2008) 103508, [\href{http://xxx.lanl.gov/abs/0709.3074}{{\tt
  arXiv:0709.3074}}].

\bibitem{Abramo:2007mp}
L.~R. Abramo and P.~Peter, {\it {K-Bounce}},  {\em JCAP} {\bf 0709} (2007) 001,
  [\href{http://xxx.lanl.gov/abs/0705.2893}{{\tt arXiv:0705.2893}}].

\bibitem{Falciano:2008gt}
F.~T. Falciano, M.~Lilley, and P.~Peter, {\it {A Classical bounce: Constraints
  and consequences}},  {\em Phys.Rev.} {\bf D77} (2008) 083513,
  [\href{http://xxx.lanl.gov/abs/0802.1196}{{\tt arXiv:0802.1196}}].

\bibitem{Linde:2009mc}
A.~Linde, V.~Mukhanov, and A.~Vikman, {\it {On adiabatic perturbations in the
  ekpyrotic scenario}},  {\em JCAP} {\bf 1002} (2010) 006,
  [\href{http://xxx.lanl.gov/abs/0912.0944}{{\tt arXiv:0912.0944}}].

\bibitem{Abramo:2009qk}
L.~R. Abramo, I.~Yasuda, and P.~Peter, {\it {Non singular bounce in modified
  gravity}},  {\em Phys.Rev.} {\bf D81} (2010) 023511,
  [\href{http://xxx.lanl.gov/abs/0910.3422}{{\tt arXiv:0910.3422}}].

\bibitem{Brandenberger:2009yt}
R.~Brandenberger, {\it {Matter Bounce in Horava-Lifshitz Cosmology}},  {\em
  Phys.Rev.} {\bf D80} (2009) 043516,
  [\href{http://xxx.lanl.gov/abs/0904.2835}{{\tt arXiv:0904.2835}}].

\bibitem{Brandenberger:2011et}
R.~H. Brandenberger, {\it {String Gas Cosmology: Progress and Problems}},  {\em
  Class.Quant.Grav.} {\bf 28} (2011) 204005,
  [\href{http://xxx.lanl.gov/abs/1105.3247}{{\tt arXiv:1105.3247}}].

\bibitem{Brandenberger:2012zb}
R.~H. Brandenberger, {\it {The Matter Bounce Alternative to Inflationary
  Cosmology}},  \href{http://xxx.lanl.gov/abs/1206.4196}{{\tt
  arXiv:1206.4196}}.

\bibitem{Cai:2012va}
Y.-F. Cai, D.~A. Easson, and R.~Brandenberger, {\it {Towards a Nonsingular
  Bouncing Cosmology}},  {\em JCAP} {\bf 1208} (2012) 020,
  [\href{http://xxx.lanl.gov/abs/1206.2382}{{\tt arXiv:1206.2382}}].

\bibitem{Cai:2013vm}
Y.-F. Cai, R.~Brandenberger, and P.~Peter, {\it {Anisotropy in a Nonsingular
  Bounce}},  \href{http://xxx.lanl.gov/abs/1301.4703}{{\tt arXiv:1301.4703}}.

\bibitem{Saaidi:2012qp}
  K.~Saaidi, H.~Sheikhahmadi and A.~H.~Mohammadi,
 ``Interacting New Agegraphic Dark Energy in a Cyclic Universe,''
  Astrophys.\ Space Sci.\  {\bf 338}, 355 (2012)
  doi:10.1007/s10509-011-0944-y
  [arXiv:1201.0275 [gr-qc]].


\bibitem{Arkani-Hamed:2015bza}
  N.~Arkani-Hamed and J.~Maldacena,
  ``Cosmological Collider Physics,''
  arXiv:1503.08043 [hep-th].


\bibitem{Dimastrogiovanni:2015pla}
  E.~Dimastrogiovanni, M.~Fasiello and M.~Kamionkowski,
  ``Imprints of Massive Primordial Fields on Large-Scale Structure,''
  JCAP {\bf 1602}, 017 (2016)
  [arXiv:1504.05993 [astro-ph.CO]].


\bibitem{Chen:2015lza}
  X.~Chen, M.~H.~Namjoo and Y.~Wang,
  ``Quantum Primordial Standard Clocks,''
  JCAP {\bf 1602}, no. 02, 013 (2016)
  [arXiv:1509.03930 [astro-ph.CO]].


\bibitem{Chen:2016cbe}
  X.~Chen, M.~H.~Namjoo and Y.~Wang,
  ``Probing the Primordial Universe using Massive Fields,''
  Int.\ J.\ Mod.\ Phys.\ D {\bf 26}, no. 01, 1740004 (2016)
  [arXiv:1601.06228 [hep-th]].


\bibitem{Lee:2016vti}
  H.~Lee, D.~Baumann and G.~L.~Pimentel,
  ``Non-Gaussianity as a Particle Detector,''
  JHEP {\bf 1612}, 040 (2016)
  [arXiv:1607.03735 [hep-th]].


\bibitem{Chen:2016qce}
  X.~Chen, M.~H.~Namjoo and Y.~Wang,
  ``A Direct Probe of the Evolutionary History of the Primordial Universe,''
  Sci.\ China Phys.\ Mech.\ Astron.\  {\bf 59}, no. 10, 101021 (2016)
  [arXiv:1608.01299 [astro-ph.CO]].


\bibitem{Meerburg:2016zdz}
  P.~D.~Meerburg, M.~Münchmeyer, J.~B.~Muñoz and X.~Chen,
  ``Prospects for Cosmological Collider Physics,''
  JCAP {\bf 1703}, no. 03, 050 (2017)
  [arXiv:1610.06559 [astro-ph.CO]].

\bibitem{Chen:2016uwp}
  X.~Chen, Y.~Wang and Z.~Z.~Xianyu,
  ``Standard Model Background of the Cosmological Collider,''
  Phys.\ Rev.\ Lett.\  {\bf 118}, no. 26, 261302 (2017)
  [arXiv:1610.06597 [hep-th]].

\bibitem{Chen:2016hrz}
  X.~Chen, Y.~Wang and Z.~Z.~Xianyu,
  ``Standard Model Mass Spectrum in Inflationary Universe,''
  JHEP {\bf 1704}, 058 (2017)
  [arXiv:1612.08122 [hep-th]].


\bibitem{An:2017hlx}
  H.~An, M.~McAneny, A.~K.~Ridgway and M.~B.~Wise,
  ``Quasi Single Field Inflation in the non-perturbative regime,''
  arXiv:1706.09971 [hep-ph].

\bibitem{An:2017rwo}
  H.~An, M.~McAneny, A.~K.~Ridgway and M.~B.~Wise,
  ``Non-Gaussian Enhancements of Galactic Halo Correlations in Quasi-Single Field Inflation,''
  arXiv:1711.02667 [hep-ph].


\bibitem{Iyer:2017qzw}
  A.~V.~Iyer, S.~Pi, Y.~Wang, Z.~Wang and S.~Zhou,
  ``Strongly Coupled Quasi-Single Field Inflation,''
  JCAP {\bf 1801}, no. 01, 041 (2018)
  [arXiv:1710.03054 [hep-th]].


\bibitem{ArmendarizPicon:1999rj}
  C.~Armendariz-Picon, T.~Damour and V.~F.~Mukhanov,
{ ``k - inflation},''
  Phys.\ Lett.\ B {\bf 458}, 209 (1999)
                [\href{http://xxx.lanl.gov/abs/hep-th/9904075}
{{\tt arXiv:hep-th/9904075}}].


\bibitem{Sheikhahmadi:2015gaa}
  H.~Sheikhahmadi, S.~Ghorbani and K.~Saaidi,
{ ``Non-local scalar fields inflationary mechanism in light of Planck $2013$},''
  Astrophys.\ Space Sci.\  {\bf 357}, no. 2, 115 (2015)
                  [\href{http://xxx.lanl.gov/abs/1502.05166}
{{\tt arXiv:1502.05166}}].


\bibitem{Bassett:2005xm}
  B.~A.~Bassett, S.~Tsujikawa and D.~Wands,
  ``Inflation dynamics and reheating,''
  Rev.\ Mod.\ Phys.\  {\bf 78}, 537 (2006)
  [astro-ph/0507632].

\bibitem{Wands:2007bd}
  D.~Wands,
  ``Multiple field inflation,''
  Lect.\ Notes Phys.\  {\bf 738}, 275 (2008)
  [astro-ph/0702187 [ASTRO-PH]].


\bibitem{Emami:2013lma}
  R.~Emami,
  ``Spectroscopy of Masses and Couplings during Inflation,''
\emph{  JCAP }{\bf 1404}, 031 (2014)
  [arXiv:1311.0184 [hep-th]].

\bibitem{Sheikhahmadi:2016wyz}
  H.~Sheikhahmadi, E.~N.~Saridakis, A.~Aghamohammadi and K.~Saaidi,
  ``Hamilton-Jacobi formalism for inflation with non-minimal derivative coupling,''
\emph{  JCAP }{\bf 1610}, no. 10, 021 (2016)
  doi:10.1088/1475-7516/2016/10/021
  [arXiv:1603.03883 [gr-qc]].



\bibitem{Sheikhahmadi:2019xkx}
  H.~Sheikhahmadi,
``Schwinger-Keldysh mechanism in extended quasi single field inflation,'' { \em Eur. Phys. J. C}  \textbf{79}, (2019) 451,
  arXiv:1901.01905 [gr-qc].

\bibitem{Alishahiha:2004eh}
M.~Alishahiha, E.~Silverstein, and D.~Tong, {\it {DBI in the sky}},  {\em
  Phys.Rev.} {\bf D70} (2004) 123505,
  [\href{http://xxx.lanl.gov/abs/hep-th/0404084}{{\tt hep-th/0404084}}].

\bibitem{Langlois:2008qf}
D.~Langlois, S.~Renaux-Petel, D.~A. Steer, and T.~Tanaka, {\it {Primordial
  perturbations and non-Gaussianities in DBI and general multi-field
  inflation}},  {\em Phys.Rev.} {\bf D78} (2008) 063523,
  [\href{http://xxx.lanl.gov/abs/0806.0336}{{\tt arXiv:0806.0336}}].

\bibitem{Langlois:2009ej}
D.~Langlois, S.~Renaux-Petel, and D.~A. Steer, {\it {Multi-field DBI inflation:
  Introducing bulk forms and revisiting the gravitational wave constraints}},
  {\em JCAP} {\bf 0904} (2009) 021,
  [\href{http://xxx.lanl.gov/abs/0902.2941}{{\tt arXiv:0902.2941}}].




\bibitem{Golovnev:2008cf}
A.~Golovnev, V.~Mukhanov, and V.~Vanchurin, {\it {Vector Inflation}},  {\em
  JCAP} {\bf 0806} (2008) 009, [\href{http://xxx.lanl.gov/abs/0802.2068}{{\tt
  arXiv:0802.2068}}].

\bibitem{Adshead:2012kp}
P.~Adshead and M.~Wyman, {\it {Chromo-Natural Inflation: Natural inflation on a
  steep potential with classical non-Abelian gauge fields}},  {\em
  Phys.Rev.Lett.} {\bf 108} (2012) 261302,
  [\href{http://xxx.lanl.gov/abs/1202.2366}{{\tt arXiv:1202.2366}}].

\bibitem{Maleknejad:2011jw}
A.~Maleknejad and M.~Sheikh-Jabbari, {\it {Gauge-flation: Inflation From
  Non-Abelian Gauge Fields}},  \href{http://xxx.lanl.gov/abs/1102.1513}{{\tt
  arXiv:1102.1513}}.

\bibitem{Maleknejad:2011sq}
A.~Maleknejad and M.~Sheikh-Jabbari, {\it {Non-Abelian Gauge Field Inflation}},
   {\em Phys.Rev.} {\bf D84} (2011) 043515,
  [\href{http://xxx.lanl.gov/abs/1102.1932}{{\tt arXiv:1102.1932}}].

\bibitem{Maleknejad:2012fw}
A.~Maleknejad, M.~Sheikh-Jabbari, and J.~Soda, {\it {Gauge Fields and
  Inflation}},  \href{http://xxx.lanl.gov/abs/1212.2921}{{\tt
  arXiv:1212.2921}}.



\bibitem{Berera:1995ie}
  A.~Berera,
 ``Warm inflation,''
  Phys.\ Rev.\ Lett.\  {\bf 75}, 3218 (1995)
  doi:10.1103/PhysRevLett.75.3218
  [astro-ph/9509049].

\bibitem{Berera:1996nv}
  A.~Berera,
 ``Thermal properties of an inflationary universe,''
  Phys.\ Rev.\ D {\bf 54}, 2519 (1996)
  doi:10.1103/PhysRevD.54.2519
  [hep-th/9601134].



\bibitem{Berera:1995wh}
  A.~Berera and L.~Z.~Fang,
  ``Thermally induced density perturbations in the inflation era,''
  Phys.\ Rev.\ Lett.\  {\bf 74}, 1912 (1995)
  doi:10.1103/PhysRevLett.74.1912
  [astro-ph/9501024].

\bibitem{Berera:1996fm}
  A.~Berera,
``Interpolating the stage of exponential expansion in the early universe: A Possible alternative with no reheating,''
  Phys.\ Rev.\ D {\bf 55}, 3346 (1997)
  doi:10.1103/PhysRevD.55.3346
  [hep-ph/9612239].



\bibitem{Berera:1998gx}
  A.~Berera, M.~Gleiser and R.~O.~Ramos,
 ``Strong dissipative behavior in quantum field theory,''
  Phys.\ Rev.\ D {\bf 58}, 123508 (1998)
  doi:10.1103/PhysRevD.58.123508
  [hep-ph/9803394].

\bibitem{Berera:1999ws}
  A.~Berera,
``Warm inflation at arbitrary adiabaticity: A Model, an existence proof for inflationary dynamics in quantum field theory,''
  Nucl.\ Phys.\ B {\bf 585}, 666 (2000)
  doi:10.1016/S0550-3213(00)00411-9
  [hep-ph/9904409].





\bibitem{Bastero-Gil:2016qru}
  M.~Bastero-Gil, A.~Berera, R.~O.~Ramos and J.~G.~Rosa,
``Warm Little Inflaton,''
  Phys.\ Rev.\ Lett.\  {\bf 117}, no. 15, 151301 (2016)
  doi:10.1103/PhysRevLett.117.151301
  [arXiv:1604.08838 [hep-ph]].

\bibitem{Berera:2018tfc}
  A.~Berera, J.~Mabillard, M.~Pieroni and R.~O.~Ramos,
``Identifying Universality in Warm Inflation,''
  JCAP {\bf 1807}, no. 07, 021 (2018)
  doi:10.1088/1475-7516/2018/07/021
  [arXiv:1803.04982 [astro-ph.CO]].


\bibitem{Yokoyama:1998ju}
J.~Yokoyama and A.~D. Linde, {\it {Is warm inflation possible?}},  {\em
  Phys.Rev.} {\bf D60} (1999) 083509,
  [\href{http://xxx.lanl.gov/abs/hep-ph/9809409}{{\tt hep-ph/9809409}}].

\bibitem{BasteroGil:2010pb}
M.~Bastero-Gil, A.~Berera, and R.~O. Ramos, {\it {Dissipation coefficients from
  scalar and fermion quantum field interactions}},  {\em JCAP} {\bf 1109}
  (2011) 033, [\href{http://xxx.lanl.gov/abs/1008.1929}{{\tt
  arXiv:1008.1929}}].

\bibitem{Bartrum:2013oka}
S.~Bartrum, A.~Berera, and J.~G. Rosa, {\it {Warming up for Planck}},
  \href{http://xxx.lanl.gov/abs/1303.3508}{{\tt arXiv:1303.3508}}.

\bibitem{Naderi:2018kre}
  M.~Naderi, A.~Aghamohammadi, A.~Refaei and H.~Sheikhahmadi,
  ``The Effects of low anisotropy on non-canonical scalar field with intermediate inflation,''
  arXiv:1809.02348 [physics.gen-ph].

\bibitem{Ghadiri:2018nok}
  Z.~Ghadiri, A.~Refaei, A.~Aghamohammadi and H.~Sheikhahmadi,
  ``Constraints on warm power-law inflation in light of Planck $2013$ and $2015$ results,''
  arXiv:1809.00165 [gr-qc].


\bibitem{Turner:1983he}
M.~S. Turner, {\it {Coherent Scalar Field Oscillations in an Expanding
  Universe}},  {\em Phys. Rev.} {\bf D28} (1983) 1243.

\bibitem{Kofman:1997yn}
L.~Kofman, A.~D. Linde, and A.~A. Starobinsky, {\it {Towards the theory of
  reheating after inflation}},  {\em Phys. Rev.} {\bf D56} (1997) 3258--3295,
  [\href{http://xxx.lanl.gov/abs/hep-ph/9704452}{{\tt hep-ph/9704452}}].


\bibitem{Mazumdar:2010sa}
A.~Mazumdar and J.~Rocher, {\it {Particle physics models of inflation and
  curvaton scenarios}},  {\em Phys. Rept.} {\bf 497} (2011) 85--215,
  [\href{http://xxx.lanl.gov/abs/1001.0993}{{\tt arXiv:1001.0993}}].

\bibitem{Finelli:1998bu}
F.~Finelli and R.~H. Brandenberger, {\it {Parametric amplification of
  gravitational fluctuations during reheating}},  {\em Phys.Rev.Lett.} {\bf 82}
  (1999) 1362--1365, [\href{http://xxx.lanl.gov/abs/hep-ph/9809490}{{\tt
  hep-ph/9809490}}].

\bibitem{Bassett:1998wg}
B.~A. Bassett, D.~I. Kaiser, and R.~Maartens, {\it {General relativistic
  preheating after inflation}},  {\em Phys.Lett.} {\bf B455} (1999) 84--89,
  [\href{http://xxx.lanl.gov/abs/hep-ph/9808404}{{\tt hep-ph/9808404}}].

\bibitem{Finelli:2000ya}
F.~Finelli and R.~H. Brandenberger, {\it {Parametric amplification of metric
  fluctuations during reheating in two field models}},  {\em Phys. Rev.} {\bf
  D62} (2000) 083502, [\href{http://xxx.lanl.gov/abs/hep-ph/0003172}{{\tt
  hep-ph/0003172}}].

\bibitem{Jedamzik:2010dq}
K.~Jedamzik, M.~Lemoine, and J.~Martin, {\it {Collapse of Small-Scale Density
  Perturbations during Preheating in Single Field Inflation}},  {\em JCAP} {\bf
  1009} (2010) 034, [\href{http://xxx.lanl.gov/abs/1002.3039}{{\tt
  arXiv:1002.3039}}].

\bibitem{Jedamzik:2010hq}
K.~Jedamzik, M.~Lemoine, and J.~Martin, {\it {Generation of gravitational waves
  during early structure formation between cosmic inflation and reheating}},
  {\em JCAP} {\bf 1004} (2010) 021,
  [\href{http://xxx.lanl.gov/abs/1002.3278}{{\tt arXiv:1002.3278}}].

\bibitem{Easther:2010mr}
R.~Easther, R.~Flauger, and J.~B. Gilmore, {\it {Delayed Reheating and the
  Breakdown of Coherent Oscillations}},  {\em JCAP} {\bf 1104} (2011) 027,
  [\href{http://xxx.lanl.gov/abs/1003.3011}{{\tt arXiv:1003.3011}}].







 \bibitem{Ade:2013ktc}
{\bf Planck Collaboration} Collaboration, P.~Ade et~al., {\it {Planck 2013
  results. I. Overview of products and scientific results}},
  \href{http://xxx.lanl.gov/abs/1303.5062}{{\tt arXiv:1303.5062}}.


\bibitem{Ade:2013uln}
{\bf Planck Collaboration} Collaboration, P.~Ade et~al., {\it {Planck 2013
  results. XXII. Constraints on inflation}},
  \href{http://xxx.lanl.gov/abs/1303.5082}{{\tt arXiv:1303.5082}}.

\bibitem{Ade:2013ydc}
{\bf Planck Collaboration} Collaboration, P.~Ade et~al., {\it {Planck 2013
  Results. XXIV. Constraints on primordial non-Gaussianity}},
  \href{http://xxx.lanl.gov/abs/1303.5084}{{\tt arXiv:1303.5084}}.

\bibitem{Tonry:2003zg}
{\bf Supernova Search Team} Collaboration, J.~L. Tonry et~al., {\it
  {Cosmological results from high-z supernovae}},  {\em Astrophys.J.} {\bf 594}
  (2003) 1--24, [\href{http://xxx.lanl.gov/abs/astro-ph/0305008}{{\tt
  astro-ph/0305008}}].

\bibitem{Riess:2004nr}
{\bf Supernova Search Team} Collaboration, A.~G. Riess et~al., {\it {Type Ia
  supernova discoveries at z>1 from the Hubble Space Telescope: Evidence for
  past deceleration and constraints on dark energy evolution}},  {\em
  Astrophys.J.} {\bf 607} (2004) 665--687,
  [\href{http://xxx.lanl.gov/abs/astro-ph/0402512}{{\tt astro-ph/0402512}}].

\bibitem{Riess:2006fw}
A.~G. Riess, L.-G. Strolger, S.~Casertano, H.~C. Ferguson, B.~Mobasher, et~al.,
  {\it {New Hubble Space Telescope Discoveries of Type Ia Supernovae at z>1:
  Narrowing Constraints on the Early Behavior of Dark Energy}},  {\em
  Astrophys.J.} {\bf 659} (2007) 98--121,
  [\href{http://xxx.lanl.gov/abs/astro-ph/0611572}{{\tt astro-ph/0611572}}].

\bibitem{Riess:2011yx}
A.~G. Riess, L.~Macri, S.~Casertano, H.~Lampeitl, H.~C. Ferguson, et~al., {\it
  {A 3\% Solution: Determination of the Hubble Constant with the Hubble Space
  Telescope and Wide Field Camera 3}},  {\em Astrophys.J.} {\bf 730} (2011)
  119, [\href{http://xxx.lanl.gov/abs/1103.2976}{{\tt arXiv:1103.2976}}].

\bibitem{AdelmanMcCarthy:2007aa}
{\bf SDSS Collaboration} Collaboration, J.~K. Adelman-McCarthy et~al., {\it
  {The Sixth Data Release of the Sloan Digital Sky Survey}},  {\em
  Astrophys.J.Suppl.} {\bf 175} (2008) 297--313,
  [\href{http://xxx.lanl.gov/abs/0707.3413}{{\tt arXiv:0707.3413}}].

\bibitem{Abazajian:2008wr}
{\bf SDSS Collaboration} Collaboration, K.~N. Abazajian et~al., {\it {The
  Seventh Data Release of the Sloan Digital Sky Survey}},  {\em
  Astrophys.J.Suppl.} {\bf 182} (2009) 543--558,
  [\href{http://xxx.lanl.gov/abs/0812.0649}{{\tt arXiv:0812.0649}}].

\bibitem{Amiaux:2012bt}
{\bf Euclid collaboration} Collaboration, J.~Amiaux et~al., {\it {Euclid
  Mission: building of a Reference Survey}},
  \href{http://xxx.lanl.gov/abs/1209.2228}{{\tt arXiv:1209.2228}}.

\bibitem{Turner:1993vb}
M.~S. Turner, M.~J. White, and J.~E. Lidsey, {\it {Tensor perturbations in
  inflationary models as a probe of cosmology}},  {\em Phys.Rev.} {\bf D48}
  (1993) 4613--4622, [\href{http://xxx.lanl.gov/abs/astro-ph/9306029}{{\tt
  astro-ph/9306029}}].

\bibitem{Maggiore:1999vm}
M.~Maggiore, {\it {Gravitational wave experiments and early universe
  cosmology}},  {\em Phys.Rept.} {\bf 331} (2000) 283--367,
  [\href{http://xxx.lanl.gov/abs/gr-qc/9909001}{{\tt gr-qc/9909001}}].

\bibitem{Kudoh:2005as}
H.~Kudoh, A.~Taruya, T.~Hiramatsu, and Y.~Himemoto, {\it {Detecting a
  gravitational-wave background with next-generation space interferometers}},
  {\em Phys.Rev.} {\bf D73} (2006) 064006,
  [\href{http://xxx.lanl.gov/abs/gr-qc/0511145}{{\tt gr-qc/0511145}}].

\bibitem{Kuroyanagi:2009br}
S.~Kuroyanagi, C.~Gordon, J.~Silk, and N.~Sugiyama, {\it {Forecast Constraints
  on Inflation from Combined CMB and Gravitational Wave Direct Detection
  Experiments}},  {\em Phys.Rev.} {\bf D81} (2010) 083524,
  [\href{http://xxx.lanl.gov/abs/0912.3683}{{\tt arXiv:0912.3683}}].

\bibitem{Kawamura:2011zz}
S.~Kawamura, M.~Ando, N.~Seto, S.~Sato, T.~Nakamura, et~al., {\it {The Japanese
  space gravitational wave antenna: DECIGO}},  {\em Class.Quant.Grav.} {\bf 28}
  (2011) 094011.

\bibitem{AmaroSeoane:2012km}
P.~Amaro-Seoane, S.~Aoudia, S.~Babak, P.~Binetruy, E.~Berti, et~al., {\it
  {eLISA: Astrophysics and cosmology in the millihertz regime}},
  \href{http://xxx.lanl.gov/abs/1201.3621}{{\tt arXiv:1201.3621}}.

\bibitem{Kuroyanagi:2013ns}
S.~Kuroyanagi, C.~Ringeval, and T.~Takahashi, {\it {Early Universe Tomography
  with CMB and Gravitational Waves}},  {\em Phys. Rev. D} {\bf 87} (2013)
  083502, [\href{http://xxx.lanl.gov/abs/1301.1778}{{\tt arXiv:1301.1778}}].

\bibitem{Dunkley:2013vu}
J.~Dunkley, E.~Calabrese, J.~Sievers, G.~Addison, N.~Battaglia, et~al., {\it
  {The Atacama Cosmology Telescope: likelihood for small-scale CMB data}},
  \href{http://xxx.lanl.gov/abs/1301.0776}{{\tt arXiv:1301.0776}}.

\bibitem{Sievers:2013wk}
J.~L. Sievers, R.~A. Hlozek, M.~R. Nolta, V.~Acquaviva, G.~E. Addison, et~al.,
  {\it {The Atacama Cosmology Telescope: Cosmological parameters from three
  seasons of data}},  \href{http://xxx.lanl.gov/abs/1301.0824}{{\tt
  arXiv:1301.0824}}.

\bibitem{Hou:2012xq}
Z.~Hou, C.~Reichardt, K.~Story, B.~Follin, R.~Keisler, et~al., {\it
  {Constraints on Cosmology from the Cosmic Microwave Background Power Spectrum
  of the 2500-square degree SPT-SZ Survey}},
  \href{http://xxx.lanl.gov/abs/1212.6267}{{\tt arXiv:1212.6267}}.

\bibitem{Story:2012wx}
K.~Story, C.~Reichardt, Z.~Hou, R.~Keisler, K.~Aird, et~al., {\it {A
  Measurement of the Cosmic Microwave Background Damping Tail from the
  2500-square-degree SPT-SZ survey}},
  \href{http://xxx.lanl.gov/abs/1210.7231}{{\tt arXiv:1210.7231}}.

\bibitem{Baumann:2008aq}
{\bf CMBPol Study Team} Collaboration, D.~Baumann et~al., {\it {CMBPol Mission
  Concept Study: Probing Inflation with CMB Polarization}},  {\em AIP
  Conf.Proc.} {\bf 1141} (2009) 10--120,
  [\href{http://xxx.lanl.gov/abs/0811.3919}{{\tt arXiv:0811.3919}}].

\bibitem{Crill:2008rd}
B.~Crill, P.~Ade, E.~Battistelli, S.~Benton, R.~Bihary, et~al., {\it {SPIDER: A
  Balloon-borne Large-scale CMB Polarimeter}},
  \href{http://xxx.lanl.gov/abs/0807.1548}{{\tt arXiv:0807.1548}}.

\bibitem{Zaldarriaga:2003du}
M.~Zaldarriaga, S.~R. Furlanetto, and L.~Hernquist, {\it {21 Centimeter
  fluctuations from cosmic gas at high redshifts}},  {\em Astrophys.J.} {\bf
  608} (2004) 622--635, [\href{http://xxx.lanl.gov/abs/astro-ph/0311514}{{\tt
  astro-ph/0311514}}].

\bibitem{Lewis:2007kz}
A.~Lewis and A.~Challinor, {\it {The 21cm angular-power spectrum from the dark
  ages}},  {\em Phys. Rev.} {\bf D76} (2007) 083005,
  [\href{http://xxx.lanl.gov/abs/astro-ph/0702600}{{\tt astro-ph/0702600}}].

\bibitem{Tegmark:2008au}
M.~Tegmark and M.~Zaldarriaga, {\it {The Fast Fourier Transform Telescope}},
  {\em Phys. Rev.} {\bf D79} (2009) 083530,
  [\href{http://xxx.lanl.gov/abs/0805.4414}{{\tt arXiv:0805.4414}}].

\bibitem{Barger:2008ii}
V.~Barger, Y.~Gao, Y.~Mao, and D.~Marfatia, {\it {Inflationary Potential from
  21 cm Tomography and Planck}},  {\em Phys. Lett.} {\bf B673} (2009) 173--178,
  [\href{http://xxx.lanl.gov/abs/0810.3337}{{\tt arXiv:0810.3337}}].

\bibitem{Mao:2008ug}
Y.~Mao, M.~Tegmark, M.~McQuinn, M.~Zaldarriaga, and O.~Zahn, {\it {How
  accurately can 21 cm tomography constrain cosmology?}},  {\em Phys. Rev.}
  {\bf D78} (2008) 023529, [\href{http://xxx.lanl.gov/abs/0802.1710}{{\tt
  arXiv:0802.1710}}].

\bibitem{Adshead:2010mc}
P.~Adshead, R.~Easther, J.~Pritchard, and A.~Loeb, {\it {Inflation and the
  Scale Dependent Spectral Index: Prospects and Strategies}},  {\em JCAP} {\bf
  1102} (2011) 021, [\href{http://xxx.lanl.gov/abs/1007.3748}{{\tt
  arXiv:1007.3748}}].

\bibitem{Clesse:2012th}
S.~Clesse, L.~Lopez-Honorez, C.~Ringeval, H.~Tashiro, and M.~H. Tytgat, {\it
  {Background reionization history from omniscopes}},  {\em Phys.Rev.} {\bf
  D86} (2012) 123506, [\href{http://xxx.lanl.gov/abs/1208.4277}{{\tt
  arXiv:1208.4277}}].




    \bibitem{Martin:2010kz}
J.~Martin and C.~Ringeval, {\it {First CMB Constraints on the Inflationary
  Reheating Temperature}},  {\em Phys. Rev.} {\bf D82} (2010) 023511,
  [\href{http://xxx.lanl.gov/abs/1004.5525}{{\tt arXiv:1004.5525}}].


\bibitem{Sheikhahmadi:2014rka}
  H.~Sheikhahmadi, A.~Aghamohammadi and K.~Saaidi,
  ``The effect of de Sitter like background on increasing the zero point budget of dark energy,''
  Adv.\ High Energy Phys.\  {\bf 2016}, 2594189 (2016)
  doi:10.1155/2016/2594189
  [arXiv:1407.0125 [gr-qc]].




\bibitem{key-1}B. Abbott et al. (LIGO Scientific Collaboration and
Virgo Collaboration), Phys. Rev. Lett. \textbf{116}, 061102 (2016).

\bibitem{key-2}A. Einstein, Sitzungsber. K. Preuss. Akad. Wiss. \textbf{1},
688 (1916).

\bibitem{key-3}B. Abbott et al. (LIGO Scientific Collaboration and
Virgo Collaboration), Phys. Rev. Lett. \textbf{116}, 241103 (2016).

\bibitem{key-4}B. Abbott et al. (LIGO Scientific Collaboration and
Virgo Collaboration), Phys. Rev. Lett. \textbf{118}, 221101 (2017).

\bibitem{key-5}B. Abbott et al. (LIGO Scientific Collaboration and
Virgo Collaboration), Phys. Rev. Lett. \textbf{119}, 141101 (2017).

\bibitem{key-6}B. Abbott et al. (LIGO Scientific Collaboration and
Virgo Collaboration), Phys. Rev. Lett. \textbf{119}, 161101 (2017).

\bibitem{key-7}B. Abbott et al. (LIGO Scientific Collaboration and
Virgo Collaboration), arXiv:1711.05578 (2017).

\bibitem{key-8}A. N. Nitz et al., Astrophys. J \textbf{872}, 2 (2019).

\bibitem{key-9}  C.~Corda,
  ``Interferometric detection of gravitational waves: the definitive test for General Relativity,''
  Int.\ J.\ Mod.\ Phys.\ D {\bf 18}, 2275 (2009)
  doi:10.1142/S0218271809015904
  [arXiv:0905.2502 [gr-qc]].

\bibitem{key-10}  C.~Corda,
  ``The future of gravitational theories in the era of the gravitational wave astronomy,''
  Int.\ J.\ Mod.\ Phys.\ D {\bf 27}, no. 05, 1850060 (2018)
  doi:10.1142/S0218271818500608
  [arXiv:1712.10318 [gr-qc]].

\bibitem{key-11}   C.~Corda,
  ``Information on the inflaton field from the spectrum of relic gravitational waves,''
  Gen.\ Rel.\ Grav.\  {\bf 42}, 1323 (2010)
  Erratum: [Gen.\ Rel.\ Grav.\  {\bf 42}, 1335 (2010)]
  doi:10.1007/s10714-009-0895-6, 10.1007/s10714-009-0917-4
  [arXiv:0909.4133 [gr-qc]].












\bibitem{Guth:1985ya}
  A.~H.~Guth and S.~Y.~Pi,
  ``The Quantum Mechanics of the Scalar Field in the New Inflationary Universe,''
  Phys.\ Rev.\ D {\bf 32}, 1899 (1985).
  doi:10.1103/PhysRevD.32.1899

\bibitem{Lyth:1984gv}
  D.~H.~Lyth,
  ``Large Scale Energy Density Perturbations and Inflation,''
  Phys.\ Rev.\ D {\bf 31}, 1792 (1985).
  doi:10.1103/PhysRevD.31.1792

\bibitem{Halliwell:1986ja}
  J.~J.~Halliwell,
  ``Scalar Fields in Cosmology with an Exponential Potential,''
  Phys.\ Lett.\ B {\bf 185}, 341 (1987).
  doi:10.1016/0370-2693(87)91011-2

\bibitem{Sasaki:1986hm}
  M.~Sasaki,
  ``Large Scale Quantum Fluctuations in the Inflationary Universe,''
  Prog.\ Theor.\ Phys.\  {\bf 76}, 1036 (1986).
  doi:10.1143/PTP.76.1036



\bibitem{Halliwell:1989vw}
  J.~J.~Halliwell,
  ``Decoherence in Quantum Cosmology,''
  Phys.\ Rev.\ D {\bf 39}, 2912 (1989).
  doi:10.1103/PhysRevD.39.2912


\bibitem{Calzetta:1990bb}
  E.~Calzetta,
  ``Anisotropy dissipation in quantum cosmology,''
  Phys.\ Rev.\ D {\bf 43}, 2498 (1991).
  doi:10.1103/PhysRevD.43.2498


\bibitem{Paz:1991ze}
  J.~P.~Paz and S.~Sinha,
  ``Decoherence and back reaction in quantum cosmology: Multidimensional minisuperspace examples,''
  Phys.\ Rev.\ D {\bf 45}, 2823 (1992).
  doi:10.1103/PhysRevD.45.2823

\bibitem{Calzetta:1993qe}
  E.~Calzetta and B.~L.~Hu,
``Noise and fluctuations in semiclassical gravity,''
  Phys.\ Rev.\ D {\bf 49}, 6636 (1994)
  doi:10.1103/PhysRevD.49.6636
  [gr-qc/9312036].


\bibitem{Matacz:1992tp}
  A.~L.~Matacz,
  ``The Coherent state representation of quantum fluctuations in the early universe,''
  Phys.\ Rev.\ D {\bf 49}, 788 (1994)
  doi:10.1103/PhysRevD.49.788
  [gr-qc/9212008].




\bibitem{Berera:1998px}
  A.~Berera, M.~Gleiser and R.~O.~Ramos,
 ``A First principles warm inflation model that solves the cosmological horizon / flatness problems,''
  Phys.\ Rev.\ Lett.\  {\bf 83}, 264 (1999)
  doi:10.1103/PhysRevLett.83.264
  [hep-ph/9809583].


\bibitem{Kubo:1965} R. Kubo, \emph{Statistical Mechanics: An Advanced Course with Problems and Solutions}, North-
Holland, Amsterdam, 1965.



\bibitem{Kubo:1957mj}
  R.~Kubo,
  ``Statistical mechanical theory of irreversible processes. 1. General theory and simple applications in magnetic and conduction problems,''
  J.\ Phys.\ Soc.\ Jap.\  {\bf 12}, 570 (1957).
  doi:10.1143/JPSJ.12.570

\bibitem{Callen:1951vq}
  H.~B.~Callen and T.~A.~Welton,
``Irreversibility and generalized noise,''
  Phys.\ Rev.\  {\bf 83}, 34 (1951).
  doi:10.1103/PhysRev.83.34





\bibitem{Caldeira:1981rx}
  A.~O.~Caldeira and A.~J.~Leggett,
``Influence of dissipation on quantum tunneling in macroscopic systems,''
  Phys.\ Rev.\ Lett.\  {\bf 46}, 211 (1981).
  doi:10.1103/PhysRevLett.46.211





\bibitem{Mota:2003tc}
  D.~F.~Mota and J.~D.~Barrow,
  ``Varying alpha in a more realistic Universe,''
  Phys.\ Lett.\ B {\bf 581}, 141 (2004)
  doi:10.1016/j.physletb.2003.12.016
  [astro-ph/0306047].

\bibitem{Khoury:2003aq}
  J.~Khoury and A.~Weltman,
``Chameleon fields: Awaiting surprises for tests of gravity in space,''
  Phys.\ Rev.\ Lett.\  {\bf 93}, 171104 (2004)
  doi:10.1103/PhysRevLett.93.171104
  [astro-ph/0309300].

\bibitem{Khoury:2003rn}
  J.~Khoury and A.~Weltman,
``Chameleon cosmology,''
  Phys.\ Rev.\ D {\bf 69}, 044026 (2004)
  doi:10.1103/PhysRevD.69.044026
  [astro-ph/0309411].

\bibitem{Brax:2004px}
  P.~Brax, C.~van de Bruck, A.~C.~Davis, J.~Khoury and A.~Weltman,
``Chameleon dark energy,''
  AIP Conf.\ Proc.\  {\bf 736}, no. 1, 105 (2004)
  doi:10.1063/1.1835177
  [astro-ph/0410103].

\bibitem{Khoury:2013yya}
  J.~Khoury,
``Chameleon Field Theories,''
  Class.\ Quant.\ Grav.\  {\bf 30}, 214004 (2013)
  doi:10.1088/0264-9381/30/21/214004
  [arXiv:1306.4326 [astro-ph.CO]].


\bibitem{Waterhouse:2006wv}
  T.~P.~Waterhouse,
  ``An Introduction to Chameleon Gravity,''
  astro-ph/0611816.



\bibitem{Clifton:2006vm}
  T.~Clifton and J.~D.~Barrow,
``Decaying gravity,''
  Phys.\ Rev.\ D {\bf 73}, 104022 (2006)
  doi:10.1103/PhysRevD.73.104022
  [gr-qc/0603116].


\bibitem{Das:2008iq}
  S.~Das and N.~Banerjee,
``Brans-Dicke Scalar Field as a Chameleon,''
  Phys.\ Rev.\ D {\bf 78}, 043512 (2008)
  doi:10.1103/PhysRevD.78.043512
  [arXiv:0803.3936 [gr-qc]].



\bibitem{Saaidi:2011zza}
  K.~Saaidi, A.~Mohammadi and H.~Sheikhahmadi,
``$\gamma$ Parameter and Solar System constraint in Chameleon Brans Dick theory,''
  Phys.\ Rev.\ D {\bf 83}, 104019 (2011)
  doi:10.1103/PhysRevD.83.104019
  [arXiv:1201.0271 [gr-qc]].

\bibitem{Saaidi:2012rj}
  K.~Saaidi, A.~Mohammadi, T.~Golanbari, H.~Sheikhahmadi and B.~Ratra,
  ``Quark-Hadron Phase Transition for a Chameleon Brans-Dicke model of Brane Gravity,''
  Phys.\ Rev.\ D {\bf 86}, 045007 (2012)
  doi:10.1103/PhysRevD.86.045007
  [arXiv:1201.0372 [gr-qc]].


\bibitem{Saaidi:2010ts}
  K.~Saaidi, H.~Sheikhahmadi and J.~Afzali,
``Chameleon mechanism with a new potential,''
  Astrophys.\ Space Sci.\  {\bf 333}, 501 (2011)
  doi:10.1007/s10509-011-0675-0
  [arXiv:1011.0075 [physics.gen-ph]].

\bibitem{Farajollahi:2010bn}
  H.~Farajollahi and A.~Salehi,
  ``Attractors, Statefinders and Observational Measurement for Chameleonic Brans--Dicke Cosmology,''
  JCAP {\bf 1011}, 006 (2010)
  doi:10.1088/1475-7516/2010/11/006
  [arXiv:1010.3589 [gr-qc]].

\bibitem{Farajollahi:2011ym}
  H.~Farajollahi, A.~Salehi, F.~Tayebi and A.~Ravanpak,
 ``Stability Analysis in Tachyonic Potential Chameleon cosmology,''
  JCAP {\bf 1105}, 017 (2011)
  doi:10.1088/1475-7516/2011/05/017
  [arXiv:1105.4045 [gr-qc]].







\bibitem{Mota:2011nh}
  D.~F.~Mota and C.~A.~O.~Schelpe,
  ``Evolution of the Chameleon Scalar Field in the Early Universe,''
  Phys.\ Rev.\ D {\bf 86}, 123002 (2012)
  doi:10.1103/PhysRevD.86.123002
  [arXiv:1108.0892 [astro-ph.CO]].

\bibitem{Hinterbichler:2013we}
  K.~Hinterbichler, J.~Khoury, H.~Nastase and R.~Rosenfeld,
 ``Chameleonic inflation,''
  JHEP {\bf 1308}, 053 (2013)
  doi:10.1007/JHEP08(2013)053
  [arXiv:1301.6756 [hep-th]].

\bibitem{Creminelli:2013nua}
  P.~Creminelli, J.~Gleyzes, L.~Hui, M.~Simonović and F.~Vernizzi,
 ``Single-Field Consistency Relations of Large Scale Structure. Part III: Test of the Equivalence Principle,''
  JCAP {\bf 1406}, 009 (2014)
  doi:10.1088/1475-7516/2014/06/009
  [arXiv:1312.6074 [astro-ph.CO]].


\bibitem{Saba:2017xur}
  N.~Saba and M.~Farhoudi,
``Chameleon Field Dynamics During Inflation,''
  Int.\ J.\ Mod.\ Phys.\ D {\bf 27}, no. 04, 1850041 (2017)
  doi:10.1142/S0218271818500414
  [arXiv:1711.09682 [gr-qc]].



\bibitem{Carroll:1998zi}
  S.~M.~Carroll,
``Quintessence and the rest of the world,''
  Phys.\ Rev.\ Lett.\  {\bf 81}, 3067 (1998)
  doi:10.1103/PhysRevLett.81.3067
  [astro-ph/9806099].


\bibitem{Damour:1994zq}
  T.~Damour and A.~M.~Polyakov,
  ``The String dilaton and a least coupling principle,''
  Nucl.\ Phys.\ B {\bf 423}, 532 (1994)
  doi:10.1016/0550-3213(94)90143-0
  [hep-th/9401069].

\bibitem{Brown:1992kc}
  J.~D.~Brown,
``Action functionals for relativistic perfect fluids,''
  Class.\ Quant.\ Grav.\  {\bf 10}, 1579 (1993)
  doi:10.1088/0264-9381/10/8/017
  [gr-qc/9304026].

\bibitem{Brown:1992bq}
  J.~D.~Brown and J.~W.~York, Jr.,
``The Microcanonical functional integral. 1. The Gravitational field,''
  Phys.\ Rev.\ D {\bf 47}, 1420 (1993)
  doi:10.1103/PhysRevD.47.1420
  [gr-qc/9209014].

\bibitem{Sotiriou:2008it}
  T.~P.~Sotiriou and V.~Faraoni,
 ``Modified gravity with R-matter couplings and (non-)geodesic motion,''
  Class.\ Quant.\ Grav.\  {\bf 25}, 205002 (2008)
  doi:10.1088/0264-9381/25/20/205002
  [arXiv:0805.1249 [gr-qc]].

\bibitem{Saaidi:2013yfa}
  K.~Saaidi, H.~Sheikhahmadi, T.~Golanbari and S.~W.~Rabiei,
  ``On the holographic dark energy in chameleon scalar-tensor cosmology,''
  Astrophys.\ Space Sci.\  {\bf 348}, 233 (2013)
  doi:10.1007/s10509-013-1491-5
  [arXiv:1404.2139 [gr-qc]].

\bibitem{Aghamohammadi:2013eja}
  A.~Aghamohammadi, K.~Saaidi, A.~Mohammadi, H.~Sheikhahmadi, T.~Golanbari and S.~W.~Rabiei,
  ``Effect of an external interaction mechanism in solving agegraphic dark energy problems,''
  Astrophys.\ Space Sci.\  {\bf 345}, no. 1, 17 (2013)
  doi:10.1007/s10509-013-1386-5
  [arXiv:1402.2608 [physics.gen-ph]].


\bibitem{Sheikhahmadi:2018aux}
  H.~Sheikhahmadi,
 ``Comments on $^"$Cosmic evolution in Brans-Dicke chameleon cosmology$^"$,''
  Eur.\ Phys.\ J.\ Plus {\bf 133}, 366 (2018)
  doi:10.1140/epjp/i2018-12235-3
  [arXiv:1802.06358 [gr-qc]].



\bibitem{Saaidi:2013pfa}
  K.~Saaidi,
  ``(Non-) geodesic motion in chameleon Brans Dicke model,''
  Astrophys.\ Space Sci.\  {\bf 345}, 431 (2013)
  doi:10.1007/s10509-013-1407-4
  [arXiv:1205.3542 [gr-qc]].






\end{thebibliography}
\end{document}